\documentclass[a4paper, 10pt]{article}

\usepackage{fullpage}
\usepackage{hyperref,graphicx,subcaption}
\usepackage{comment}
\usepackage{amssymb}
\usepackage[T1]{fontenc}
\usepackage[utf8]{inputenc}
\usepackage{combelow}
\usepackage{multirow}

\usepackage{calc}

\begin{document}

\title{Network analytics for drug repurposing in COVID-19}

\author{

	Nicoleta Siminea$^{1,2}$ \and 
	Victor Popescu$^3$ \and
	Jose Angel Sanchez Martin$^4$ \and

    Daniela Florea$^1$ \and
    Georgiana Gavril$^1$ \and 
    Ana-Maria Gheorghe$^1$ \and
    Corina I\cb{t}cu\cb{s}$^1$ \and
    Krishna Kanhaiya$^3$ \and
    Octavian Pacioglu$^1$ \and
    Laura Ioana Popa$^1$ \and
    Romica Trandafir$^1$ \and
    Maria Iris Tu\cb{s}a$^1$ \and
    
    Manuela Sidoroff$^1$ \and
    Mihaela P\u{a}un$^{1,5}$ \and
    Eugen Czeizler$^{3,1}$ \and
    Andrei P\u{a}un$^{1,2}$ \and
    Ion Petre$^{6,1,*}$

}

\footnotetext[1]{
	National Institute of Research and Development for Biological Sciences, Bucharest, Romania
}

\footnotetext[2]{
	Faculty of Mathematics and Computer Science, University of Bucharest, Bucharest, Romania
}

\footnotetext[3]{
	Department of Information Technologies, \r{A}bo Akademi University, Turku, Finland
}

\footnotetext[4]{
     Department of Computer Science, Technical University of Madrid, Spain
}

\footnotetext[5]{
	Faculty of Administration and Business, University of Bucharest, Bucharest, Romania
}

\footnotetext[6]{
    Department of Mathematics and Statistics, University of Turku, Finland
}

\renewcommand*{\thefootnote}{\fnsymbol{footnote}}

\footnotetext[1]{
    Address for correspondence: \texttt{ion.petre@utu.fi}
}

\renewcommand*{\thefootnote}{\arabic{footnote}}

\date{}

\maketitle

\begin{abstract}

    To better understand the potential of drug repurposing in COVID-19, we analyzed control strategies over essential host factors for SARS-CoV-2 infection. We constructed comprehensive directed protein-protein interaction networks integrating the top ranked host factors, drug target proteins, and directed protein-protein interaction data. We analyzed the networks to identify drug targets and combinations thereof that offer efficient control over the host factors. We validated our findings against clinical studies data and bioinformatics studies. Our method offers a new insight into the molecular details of the disease and into  potentially new therapy targets for it. Our approach for drug repurposing is significant beyond COVID-19 and may be applied also to other diseases. 


\end{abstract}

\section{Introduction}
\label{section-introduction}

The COVID-19 pandemic has caused more than 159 million infections worldwide, with more than 3.3 million deaths (as of May 2021) \cite{Dong:2020te}. Several vaccines are available since the fall 2020 and this has helped curtail the infection rate worldwide. Many drugs are investigated in clinical trials, and several have been approved or recommended for use, including remdesivir, dexamethasone, and some combinations of monoclonal antibodies. It remains of major interest to gain a system-level understanding of the molecular details of the disease and to translate them into effective treatment strategies for the disease. Such data is increasingly available, for example on the human proteins that associate with SARS-CoV-2 proteins upon infection \cite{gordon2020sars} and the host genes critical for the infection with SARS-CoV-2 \cite{daniloski2021identification}. Computational methods can integrate such data into comprehensive models to help identify promising drugs and drug combination candidates \cite{Tomazou:2021uk}.

\begin{figure}[htb]
    \begin{center}
        \includegraphics[width=0.9\textwidth]{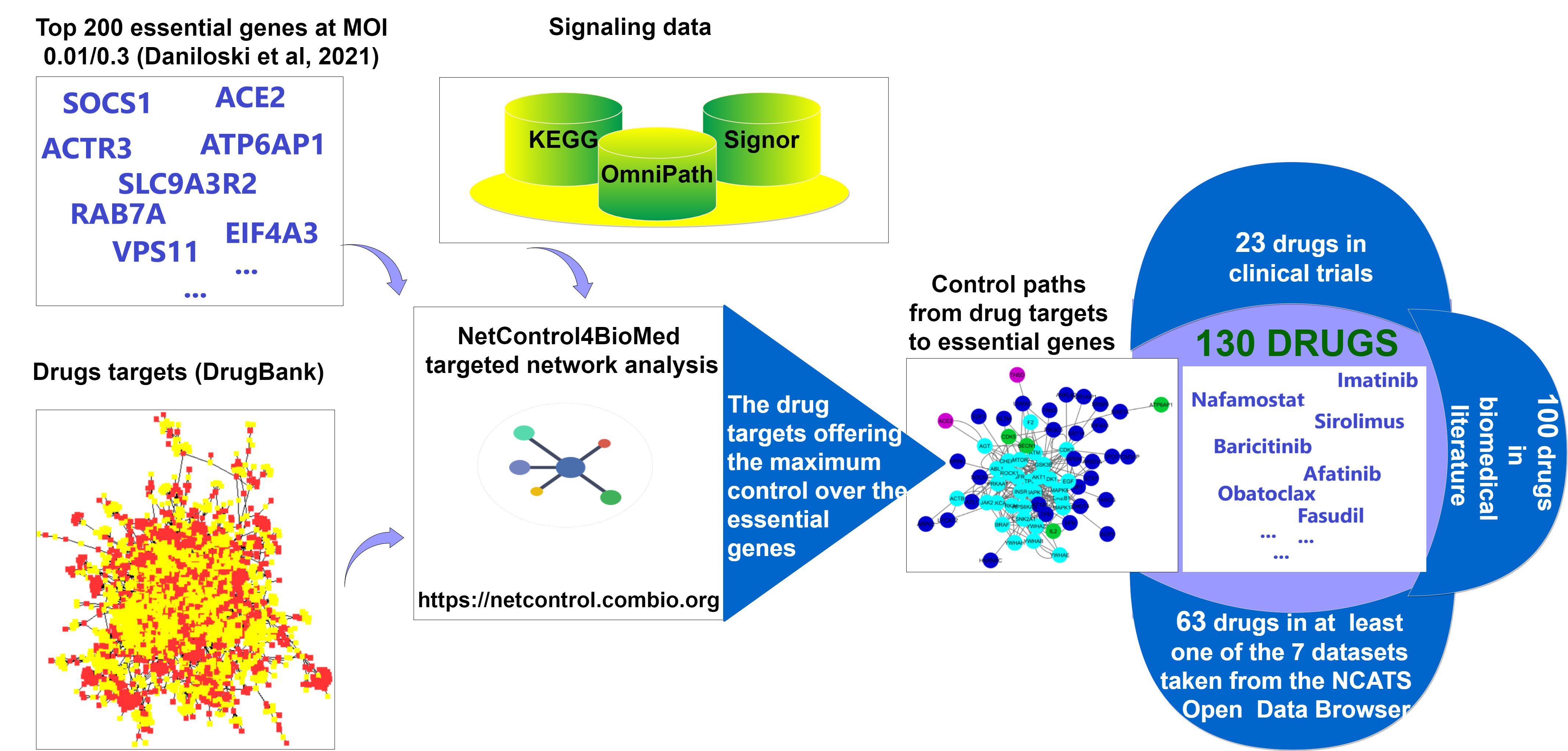}
    \end{center}
    \caption{Network controllability for drug repurposing: study design. We included all the approved and investigational small molecule drugs, except for those illicit or nutraceutical. For each drug (red in the bottom-left network) we identified their drug targets (\cite{drugbank}) (yellow in the bottom-left network). We also included the top 200 host factors required for SARS-CoV-2 infection found by \cite{daniloski2021identification} for viral loads MOI 0.01 and MOI 0.3, and all protein-protein directed interactions from KEGG, OmniPath and Signor. Using the \emph{NetControl4BioMed} platform (\cite{netcontrol4biomed2021}) we identified all the control paths of length at most 3 from drug targets to the host factors. We ranked the drugs based on the number of host factors they can control through any of their targets.}
    \label{figure-workflow}
\end{figure}

The recent results of \cite{daniloski2021identification} offer a dataset with a strong therapeutical potential. Through a number of genome-scale loss-of-function screens, they identified and ranked host factors required for the SARS-CoV-2 viral infection, i.e., genes whose loss of function confers resistance to the infection. They identified the relevant host factors both at a lower viral load of MOI 0.01 and a higher one of MOI 0.3. Out of the 200 top ranked genes, only a small fraction are drug targetable \cite{drugbank} (24 from the MOI 0.01 host factors and 23 from the MOI 0.3 ones). We asked whether more of the host factors can be targeted through drugs acting upstream of them. To investigate this question we built two directed protein-protein interaction networks (one for each MOI dataset) integrating interaction data upstream of the essential SARS-CoV-2 host factors (separately for MOI 0.01 and MOI 0.3) and interaction data downstream of all currently available drug targets from DrugBank \cite{drugbank}. We analyzed the networks in the framework of control theory \cite{liu:2011} and sought to identify minimal combinations of drug targets that offer control (through cascading signals in the interaction network) over the essential host factors (Figure \ref{figure-workflow}). We focused on short control paths from drug targets to host factors to minimize the possible dissipation of the drug's influence along the path. Control here is understood in the sense of structural network controllability and its results offer a systemic view on how to influence the host factors simultaneously through available drug targets. Moreover, the resulting drug targets (and the drugs acting on them) can be ranked with respect to how many host factors each can control in the network, independently of the other drug targets. Combining these results offers a number of drugs and drug combinations that are potentially efficient at influencing the host factors. The results of structural controllability are qualitative: they offer therapeutically promising drugs and drug combinations, but do not offer numerical indications on the optimal concentrations and possible toxicities. We compared our findings with clinical trials data and bioinformatics analyses. We found a number of drugs that have been investigated in clinical trials, but also some new ones, not yet studied in connection to COVID-19. Our key advance, therefore, is a new system-level insight into the molecular details of the disease that is able to offer a significant number of potentially efficient therapies based on drug repurposing.

\section{Results}
\label{section-results}

\subsection{COVID-19--specific directed protein-protein interaction networks}
\label{section-results-ppi-networks}

We constructed directed protein-protein interaction (PPI) networks around the host factors identified in \cite{daniloski2021identification} to be required for SARS-CoV-2 infection. We constructed two different interaction networks, one for each set of top-ranked host factors reported by \cite{daniloski2021identification}, one obtained at a low (0.01) multiplicity of infection (MOI), the other at a high (0.3) MOI. From each experiment we included the top 200 ranked genes, whose loss-of-function mutations led to enrichment in their pools. We also included all drug targets from DrugBank \cite{drugbank}. For the interaction data we used KEGG \cite{kanehisa:2000aa}, OmniPath \cite{turei:2016aa} and SIGNOR \cite{10.1093/nar/gkz949}. We only included the directed protein-protein interactions  found in these databases. A description of how the networks were generated is in the supplementary information, Section~\ref{section-supplementary-information-network} (Figure~\ref{figure-networks} for an illustration and Table~\ref{table-networks-statistics} for an overview of their topological properties). They were well connected (a single connected component in one, two in the other), compact (diameter equal to 10), and rich in interactions (over 20,000 interactions). Each network included more than 1000 drug targets and about one third of the top 200 host factors (70 for MOI 0.01 and 62 for MOI 0.3; the others were further from the drug targets in our networks).

We analyzed the topology of our networks using several centrality measures: degree (total degree, in-degree, and out-degree) centrality, closeness-, betweenness-, and harmonic-centrality \cite{10.1093/bib/bbz011}. We identified the top $3$ ranked proteins in each network, based on each of these centrality measures, with the results collected in the supplementary information, Table~\ref{table-networks-centrality}. The top-ranked proteins were virtually identical for both networks, with only one difference in the proteins with the highest out-degree: GSK3B was ranked third in the MOI 0.01 network, while CDK1 was similarly ranked in the MOI 0.3 network.

The proteins with the highest total- and out- degree were PRKCA, SRC and GSK3B, out of which SRC and GSK3B have been shown to be potential targets for COVID-19 (\cite{pmid32778962}). Another protein of interest is TP53, which was top 2-ranked in all the other centrality measures. TP53 is known to be active in down-regulating the SARS-CoV replication (\cite{Ma-LauerE5192}). Other highly ranked central proteins include AKT1, shown to be of significance to SARS-CoV-2 (\cite{Appelberg2020}), CDK1, identified as potential therapeutic option in for COVID-19 (\cite{Gargouri2021}), and CTNNB1, linked to co-morbidities associated with COVID-19 severity (\cite{Dolan2020}).

\subsection{Network-based identification of repurposable drug targets}
\label{section-results-drug-repurposing}

We performed target controllability analysis on the interaction networks, considering as control targets the MOI-specific top 200 host factors of \cite{daniloski2021identification}. We used as preferred control inputs the drug-targetable genes from DrugBank \cite{drugbank}. In each analysis we identified the drug(s) whose targets control the most host factors. The analysis is based on a stochastic search for paths from drug targets to host factors and so, repeated analyses identified multiple results on the same network. For each network, we repeated the analysis until three consecutive runs identified no new drugs. The analyses were run on the \emph{NetControl4BioMed} platform (\cite{netcontrol4biomed2021}).

We found that the host factors can be controlled though 35 drug-targets for the MOI 0.01 network and through 15 for the MOI 0.3 network (see Figure~\ref{figure-correspondence-drug-target-essential} in the supplementary information), of which 10 are common for both networks: ACTB, AKT1, ATM, ATP6AP1, CSNK2A1, CDK2, EGF, MAPK14, MTOR and TP53.  Of these 10, two are essential host factors for the SARS-CoV-2 infection \cite{daniloski2021identification}: ACTB (in the MOI 0.3 experiment) and ATP6AP1 (in both MOI experiments). Additionally, ATP6AP1 is also known to interact directly with SARS-CoV-2 \cite{gordon2020sars}. We identified $116$ drugs from the analysis of the MOI 0.01 network acting on these targets and $55$ drugs from the analysis of the MOI 0.3 network, with $41$ common to both of them (see Figure~\ref{figure-categories} in the supplementary information for the different categories of drugs).

\begin{figure}[htb]
    \begin{center}
        \includegraphics[width=\textwidth]{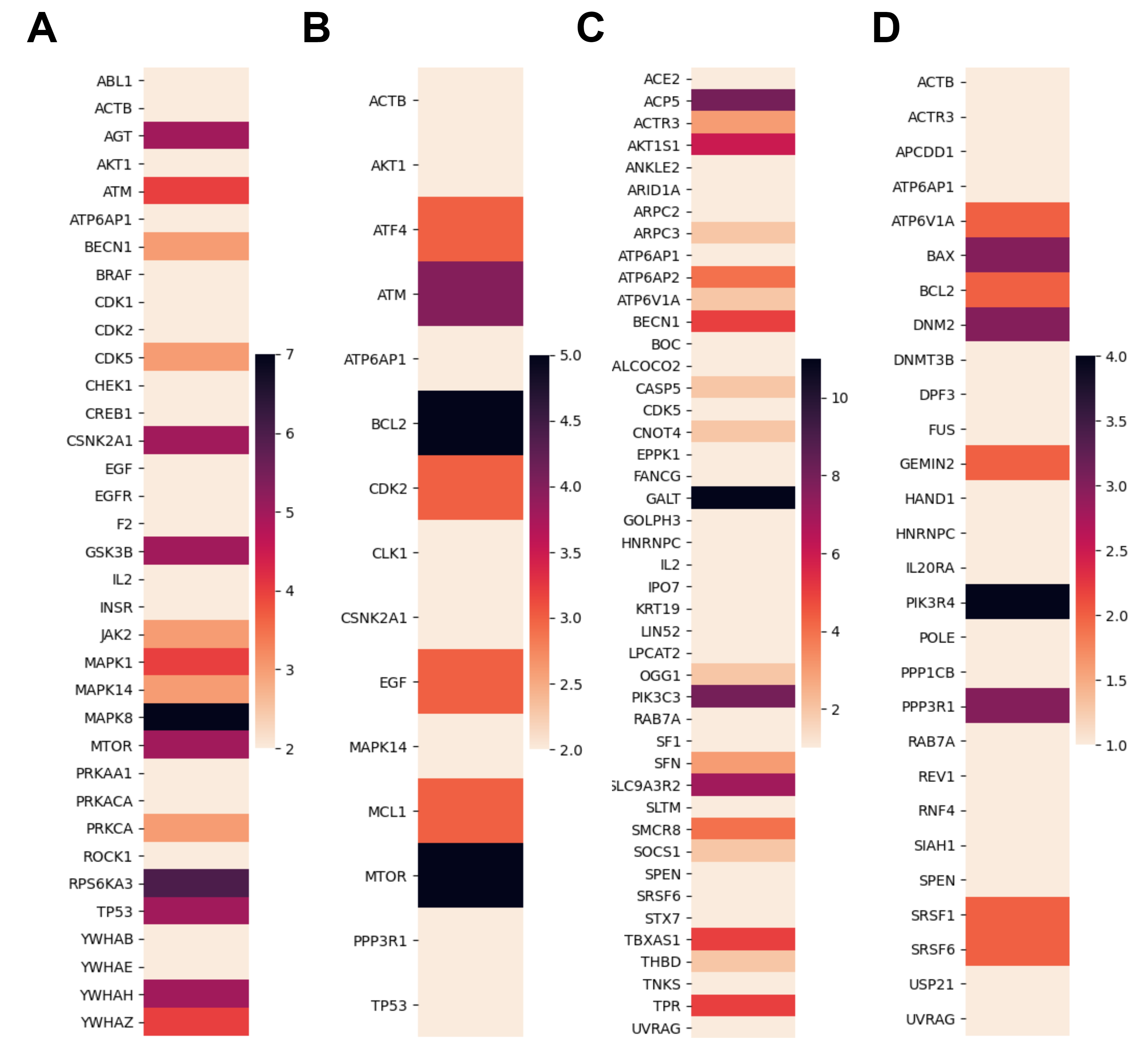}
    \end{center}
    \caption{{\textbf{A-B:} Drug-targets and the number of host factors they can control (\textbf{A}: MOI 0.01, \textbf{B}: MOI 0.3). \textbf{C-D:} Host factors for the SARS-CoV-2 infection} and  the number of drug-target genes that can be used to control them (\textbf{C}: MOI 0.01, \textbf{D}: MOI 0.3).}
    \label{figure-essential-genes-and-drug-targets}
\end{figure}

We found that not all host factors can be controlled by repurposed drugs: only 44 MOI 0.01 host factors and only 28 MOI 0.3 host factors were included in the control results found by our analysis (see Figure~\ref{figure-correspondence-drug-target-essential} in the supplementary information). In the MOI 0.01 analysis the gene that turned out to be the easiest to control is GALT: it can be controlled by more than 10 drug-targets (Figure~\ref{figure-essential-genes-and-drug-targets}). By inhibiting GALT, the galactose levels may increase (\cite{uniprot2015uniprot}), and SARS-CoV-2 does not replicate in the presence of galactose (\cite{Codo2020}). On the other hand, the well-known ACE2 is one of the hardest to target: it can be controlled by only one drug-target, namely AGT.

Our analysis also identified the control paths from the drug targets to the host factors. The control paths were identical for the MOI 0.01 and the MOI 0.3 data in several cases: ATP6AP1 (which is a host factor on both lists, as well as a drug target), RAB7A (controllable by ATP6AP1 through the direct interaction ATP6AP1 $\rightarrow$ RAB7A), UVRAG (controllable by MTOR through MTOR $\rightarrow$ UVRAG), SRSF6 (controllable by EGF through EGF $\rightarrow$ CLK1 $\rightarrow$ SRPK1 $\rightarrow$ SRSF6), and ATP6V1A (controllable by both CSNK2A1 and MTOR through CSNK2A1 $\rightarrow$ YY1 $\rightarrow$ ATP6V1A, MTOR $\rightarrow$ YY1 $\rightarrow$ ATP6V1A), ACTR3 (controllable by ACTB through ACTB $\rightarrow$ CTTN $\rightarrow$ ACTR3). Three of these  interact with SARS-CoV-2: ATP6AP1 with nsp6, ATP6V1A with M, and RAB7A with nsp7 \cite{gordon2020sars}, while several others are relevant for other viruses (UVRAG, SRSF6, ATP6AP1, RAB7A, ATP6V1A, ACTR3).

A ranking of the theoretical efficacy of a drug target can be done on the basis of how many host factors was found to control in our analyses (Figure~\ref{figure-essential-genes-and-drug-targets}A-B). Also, a ranking of how easily controllable a host factor is can be done on the basis of how many drug targets can control it (Figure~\ref{figure-essential-genes-and-drug-targets}C-D). For the MOI 0.01 host factors, the top-ranked drug target was MAPK8, which controls 7 host factors. Interestingly, MAPK8 was not identified as a control input in our analyses for the MOI 0.3 list of host factors. For MAPK8 we found two drugs acting on it: minocycline and tamoxifen. The top-ranked drug targets occurring in both sets of results is MTOR, controlling 5 host factors from each list, and ATM, controlling 4 host factors from each list. There are several drugs that act on MTOR: everolimus, pimecrolimus, ridaforolimus, rimiducid, SF1126, sirolimus, temsirolimus and XL765. For the 0.3 MOI list of host factors, the top-ranked drug target, in addition to MTOR, is BCL2, that regulates cell death and attenuates inflammation (\cite{uniprot2015uniprot}). The following drugs act on this target: apoptone, dexibuprofen, docetaxel, eribulin, ibuprofen, isosorbide, navitoclax, obatoclax, paclitaxel, paclitaxel docosahexaenoic acid, rasagiline and venetoclax.

\subsection{Potentially repurposable drugs}
\label{section-results-repurposable-drugs}

Using data from DrugBank (\cite{drugbank}) we found 130 drugs acting on the drug targets resulting from the two MOI datasets and their corresponding control analyses, including 41 that act on the 10 drug-targets common to the two analyses. The results are listed in the supplementary information (Table~\ref{table-networks-statistics}). Of these drugs, 72 are approved for at least one condition according to \cite{drugbank}. Most are drugs used in oncology, but we have also found direct inhibitors of thrombin, anti-inflammatory, estrogens, and mood-stabilizers. In terms of their cellular location of action, most drug targets are located in the surrounding of cytoplasmic vesicles, and in the cell-substrate junction between the cell and extracellular matrix (see Figure~\ref{figure-localization} in the supplementary information).

\subsubsection*{Antineoplastic and immunomodulating agents.}

Most drugs identified in our analyses are antineoplastic agents and, within this group, most of them are protein kinase inhibitors. Some of the drugs we identified are not yet approved, and therefore do not have ATC codes (e.g. alvocidib, gensitein, pelitinib, seliciclib). The approved drugs are summarized in Table~\ref{table-protein-kinase-inhibitors}, grouped by their targets in our networks. Some of these drugs were included in this list through some of their secondary targets (e.g. brigatinib on INSR, and dasatinib on MAPK14), and other drugs are included through several of their targets, potentially indicating increased efficacy (e.g. bosutinib and brigatinib).

\begin{table}[htb]
    \caption{The top-ranked antineoplastic and immunomodulating agents identified in our analyses and their targets.}
    \label{table-protein-kinase-inhibitors}
    \begin{center}
        \begin{tabular}{c|l}
            {\textbf{Control inputs
            }} & {\textbf{Drugs targeting  the control inputs}}\\
            \hline
            {ABL1} & bosutinib, brigatinib, dasatinib, imatinib, nilotinib, ponatinib, regorafenib \\
            \hline
            {BRAF} & dabrafenib, encorafenib, regorafenib, ripretinib, sorafenib, vemurafenib \\
            \hline
            CDK2 & bosutinib\\
            \hline
            \multirow{2}{*}{EGFR} & afatinib, brigatinib, dacomitinib, erlotinib, gefitinib, icotinib, lapatinib, \\& neratinib, olmutinib, osimertinib, vandetanib, zanubrutinib \\
            \hline
            INSR & brigatinib\\
            \hline
            {JAK2} & entrectinib, fedratinib, ruxolitinib, zanubrutinib \\
            \hline
            MAPK14 & dasatinib \\
            \hline
            MTOR & everolimus, ridaforolimus, temsirolimus \\
            \hline
            PRKCA & midostaurin \\
        \end{tabular}
    \end{center}
\end{table}

Some of these inhibitors have been investigated in connection with several viruses, including SARS-CoV-2. The EGFR inhibitors, which are principally used in non-small cell lung cancers or breast cancers, may act on SARS-CoV-2 virus replication \cite{klann2020growth}, while JAK2 inhibitors could act on SARS-CoV-2 cytokine storm because IL-6 and GM-CSF, which are stimulated in this infection, depend on JAK2 signaling \cite{allegra2020immunopathology}. The mTOR pathway can be targeted by many viruses (eg IAV, MERS) to promote their replication. Its inhibition was shown to lead to a decrease in SARS-CoV-2 virus production \cite{Appelberg2020}.

Other antineoplastic agents found in our analyses, docetaxel, paclitaxel, eribulin, venetoclax, are not protein kinase inhibitors. They were identified in the network corresponding to the MOI 0.3 experiment.

We also obtained selective immunosuppressants such as: baricitinib, si\-ro\-li\-mus, tofacitinib, and a calcineurin inhibitor, voclosporin. Sirolimus, a drug used to prevent organ rejection in renal transplants and suggested for COVID-19 also in \cite {romanelli2020sirolimus}, was identified in both network analyses, due to its inhibitory effect on MTOR, while baricitinib and tofacitinib were identified only in the MOI 0.01 network analysis, due to their inhibitory action on JAK2. Both are used in rheumatoid arthritis, and baricitinib has FDA approval for use in COVID-19. We obtained voclosporin only for the MOI 0.3 network because of its inhibitory effect on PPP3R1 subunit of calcineurin, which leads to blocking the transcription of early inflammatory cytokines (\cite{drugbank}).

\subsubsection*{Antithrombotic agents}

There is a known link between COVID-19 and coagulopathy: in many severe cases, disseminated intravascular coagulation is observed  \cite{giannis2020coagulation}. In COVID-19, there is an activation of the coagulation cascade which leads to an increase in the amounts of fibrin that are deposited in various organs \cite{costello2019disseminated}, which is associated with higher mortality \cite{connors2020thromboinflammation}. Our analysis revealed several compounds in use as antithrombotic agents (argatroban, bivalirudin, dabigatran etexilate and ximelagatran), all based on their inhibitory effect on prothrombin. We also obtained investigational drugs (flovagatran, gabexate, nafamostat), and an F2 agonist, in other words an antihemorrhagic (kappadione). Kappadione is a vitamin K derivative, and vitamin K has an important role in activating both pro- and anti-clotting factors in the liver, and extra-hepatic vitamin K deficiency has been observed in COVID-19 patients \cite{janssen2020vitamin}.

One drug present on our list that cannot be used in patients with COVID-19 is proflavine because it has only topical use as a disinfectant, and it is toxic and carcinogenic in mammals \cite{drugbank}. It ends up being included in our results through its targeted action on F2, that was found to control several host factors.

\subsubsection*{Estrogens}

Estrogens being included in our list is consistent with COVID-19 mortality being higher in the elderly, but also in men compared to women \cite{yang2020clinical}. One cause of these differences may be estradiol, several of which are included in our results. Estradiol regulates several pathways in the immune system \cite{kovats2015estrogen}. Our analysis revealed BECN1 as being relevant, a gene that plays a role in autophagy and may also play a role in antiviral host defense \cite{drugbank}. Using these drugs carries the risk of thromboembolism, even if they may increase the expression/activity of ACE2 in the adipose tissue and in the kidney \cite{la2020sex}. Some studies recommend estradiol for further investigation: 68,466 cases were analysed in \cite{seeland2020evidence} with the results indicating that estradiol  decreased COVID-19 fatality. However, estradiol has multiple functions in the body, so its potential adoption in COVID-19 should be cautiously verified further. 

\subsubsection*{Other compounds}

There are several natural compounds in our list such as resveratrol, caffeine, genistein, ellagic acid, alvocidib, quercetin, emodin. Of these, alvocidib, quer\-ce\-tin, resveratrol and genistein contains flavonoids. There are some antiallergic, anti-inflammatory, antiviral, antioxidant activities reported for some flavonoids \cite{middleton1998effect}, but also antithrombogenic effects because they can increase fibrinolysis \cite{nijveldt2001flavonoids}.

We also obtained lithium-based mood stabilizers as GSK3B inhibitors (lithium carbonate and lithium citrate). Targeting GSK3B is also tideglusib, a drug investigated for use in Alzheimer's disease.

We also obtained some investigational compounds that act on MAPK14: KC706, talmapimod, PH-797804, VX-702. Inhibition of p38/MAPK signaling is beneficial in SARS, DENV and IAV, and in the case of SARS-CoV-2 it suppresses cytokine production and affects viral replication \cite{bouhaddou2020global}. PH-797804 was also discussed in \cite{gilroy2020treating} for potential use to limit lung damage.

Isoprenaline is a heart stimulant, used especially in emergencies, in heart shock, but can also be used by inhalation in asthma and chronic bronchitis. In our results it came up through its inhibitory action on MAPK1.

Our analysis identified some compounds that should not be recommended for use in practice. Antibiotics are usually not recommended in a viral infection. We obtained minocycline, a drug from the class of tetracycline that is known to be preferred to macrolides because its representatives are less toxic. Another compound that is difficult to use is cefazolin, a cephalosporin that acts on the Il2 target. Invalid compounds are also those that act as a topical cream (for example, ingenol mebutate) or as eye drops (for example, netarsudil). Sucralfate is another false positive because it does not act systemically, but could help to prevent further transmission from the stomach to the intestine.

\subsection{Drug combinations}
\label{section-results-drug-combinations}

We used the results of the target controllability analyses to identify potential drug combinations (supplementary information, Table~\ref{table-networks-centrality}). For each analysis, we identified the combinations of two and three drugs whose drug targets control together the highest number of host factors. We only considered the combinations of drugs whose sets of drug targets don't fully overlap (i.e., each drug has at least one specific drug target not shared with the other drugs).

In the case of the MOI 0.01 network, we found 23 unique combinations of two drugs with a maximum number of controlled host factors. Some of them included aspirin and one of the drugs acting on MTOR (everolimus, pimecrolimus, ridaforolimus, rimiducid, SF1126, sirolimus, temsirolimus, XL765). Other combinations of aspirin are those with dasatinib, alvocidib, and phenethyl isothiocyanate. Other drug combinations are centered on alvocidib, a drug investigated for use in non-small lung cancer. Its combinations are with enzastaurin, minocycline, perifosine, tamoxifen, phenethyl isothiocyanate and quercetin. Enzastaurin and perifosine are AKT1 inhibitors and may be associated with CDK4/6 inhibitors (two of the alvocidib targets).

In the case of the MOI 0.3 network, we found 42 combinations of two drugs  with a maximum number of controlled host factors. Dasatinib and ellagic acid are used in multidrug-resistant tumors (\cite{Sachs2019}), while dasatinib plus quercetin has been shown to be useful in relieving intestinal senescence and inflammation (\cite{Saccon2021}). Other combinations obtained are those with bosutinib and any of the drugs that act on BCL2. Our analysis also gave BCL2 inhibitors associated with SF1126 or sirolimus (both MTOR inhibitors), and this combination of BCL2 with MTOR inhibitors is used in resistant acute lymphoblastic leukemia  (\cite{Iacovelli2015}). The other drug combinations include tesevatinib with MTOR inhibitors or caffeine.

For the MOI 0.01 network, we found 95 three-drug combinations  with a maximum number of controlled host factors, almost all centered on alvocidib. The fact that alvocidib acts on several targets, quite different from other drugs, explains why we got it in a large number of combinations. For the MOI 0.3 network, we found 69 three-drug combinations with a maximum number of controlled host factors, almost all centered on BCL2 inhibitors.

\subsection{Experimental and clinical validation}
\label{section-results-experimental-and-clinical-validation}

\paragraph{Validation using the NCATS OpenData|COVID-19 Portal.} Our first approach to validate the results was to search for them on the OpenData Portal \cite{brimacombe2020opendata}.  This data platform collects validation data on potential COVID-19 drugs in terms of viral entry, viral replication, in vitro infectivity, life virus infectivity and human cell toxicity. The platform lists both complete and incomplete results, e.g., untested drugs, or drugs with parts of the test results not yet available. We only focused on the results on SARS-CoV-2, and excluded the analyzes made on MERS and SARS. We also excluded human cell toxicology tests because we considered these types of tests more as a selection measure between two drugs with similar effects and not as a validation of activity because a drug can be inactive and can be very toxic or, conversely, can be well tolerated.  So, we included the Spike-ACE2 protein-protein interaction assay and counter-assay, ACE2, TMPRSS2 and 3CLpro enzymatic activity tests and SARS-CoV-2 cytophatic effect and counter assay.

\begin{figure}[htb]
    \begin{center}
        \includegraphics[width=\textwidth]{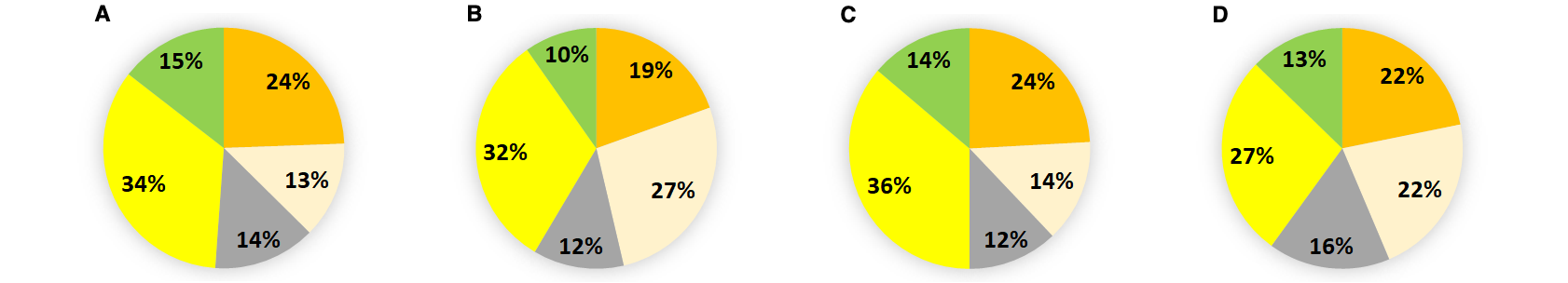}
    \end{center}
    \caption{\textbf{Drug classification based on their activity according to NCATS OpenData|COVID-19 Portal \cite{brimacombe2020opendata}}.  \textbf{A}: drugs identified in either of the two MOI networks, \textbf{B}: drugs identified in both MOI networks, \textbf{C}: drugs identified in the MOI 0.01 network, \textbf{D}: drugs identified in the MOI 0.3 network. \textbf{Color code}: green -- active, orange -- not tested, cream -- inactive, but not tested in all assays, gray -- inactive, yellow -- active in the counter assay.}
   \label{figure-drug-activity}
\end{figure}

We divided our results into the five categories used in the portal. The first category (in green in the Supplementary Table 1) is that of active drugs in at least one of the following 5 tests: Spike-ACE2 protein-protein interaction, SARS-CoV-2 cytophatic effect and enzymatic activity of ACE2, TMPRSS2 and 3CLpro. We also set the condition that, if it is active in a test, it should be inactive in the corresponding counter-test. The second category (in yellow in the Supplementary Table 1) contains active drugs in a counter assay, regardless of its status in the corresponding assay. The third category (in gray in the Supplementary Table 1) contains drugs that are inactive in all tests. The fourth (in orange in the Supplementary Table 1) includes drugs that haven't been tested at all. Finally, the fifth category (in cream in the Supplementary Table 1) contains partially tested drugs that have been found inactive in the tests performed. We illustrated the distribution of our results among these categories in Figure~\ref{figure-drug-activity}. The largest proportion is that of the drugs active in counter-assays.

We found the following results in the ``active drug'' category: afatinib, baricitinib, docetaxel, entrectinib, enzastaurin, erlotinib, fasudil, gabexate, imatinib, minocycline, nafamostat, navitoclax, ponatinib, proflavine, R-1487, ruxolitinib, sirolimus, suramin and venetoclax.

We also found several drugs active both in the assay, as well as in the counter-assay for the Spike-ACE2 protein-protein interaction, and several drugs in the counter assay for cytopathic effect (Figure~\ref{figure-overlap} in the supplementary information).

We also investigated the results on the enzymatic activity of ACE2, TMPRSS2 and 3CLpro (Figure~\ref{figure-enzymatic-activity} in the supplementary information) and found two drugs that act on ACE2 and TMPRSS2, in other words on viral entry, namely ruxolitinib and fasudil. Ruxolitinib was obtained in our study on the MOI 0.01 network as a JAK2 inhibitor. It has been previously recommended for investigation in COVID-19 (\cite{giudice2020combination}) and it has been included in clinical trials with little success (\cite{novartis_2020}). Fasudil was also obtained in the MOI 0.01 network, targeting ROCK1 and PRKACA. It may help in the associated acute lung injury and acute respiratory distress syndrome (\cite{abedi2020plausibility}). It could also up-regulate ACE2 expression and down-regulate ACE expression (\cite{ghafouri2020angiotensin}). In addition to its protective effect against lung damage, it possesses antifibrotic properties and the ability to up-regulate ACE2 (\cite{gouda2020snake}). We also found (in the MOI 0.3 network) a drug that acts on the enzymatic activity of ACE2 and 3CLpro: venetoclax, a BCL2 inhibitor.

\paragraph{In vitro validation.} About a quarter of the drugs we identified were already reported in the experimental literature. One of them is nafamostat, which inhibited mediated entry into host cells with an efficiency approximately 15 times higher than camostat mesylate (\cite {hoffmann2020nafamostat}), and also blocked infection of Calu-3 cells with an effective concentration (EC) 50 around 10 nM, while a significantly higher dose (EC50 around 30 $\mu$M) was required for VeroE6/TMPRSS2  (\cite{yamamoto2020anticoagulant}). It was the most potent drug for human lung cells (IC50 = 0.0022 $\mu$M), of the 24 drugs selected  (\cite{ko2021comparative}). In a small series of 11 cases, nafamostat mesylate therapy in combination with favipiravir could block the entry and replication of the virus as well as inhibit hypercoagulopathy  (\cite{doi2020nafamostat}).

Other drugs were tested on several cell types and were found to be active. For example, nilotinib inhibits SARS-CoV-2 in Vero‐E6 cells and Calu‐3 cells (\cite {cagno2020tyrosine}). Another drug is obatoclax, which can inhibit SARS-Cov-2 replication in vitro, in human nasal epithelial cells (\cite{varghese2021berberine}), and in Vero-E6 cells (\cite{ianevski2020potential}). Tamoxifen showed a 100-fold reduction in viral load in human cells in vitro (\cite{daniloski2021identification}), but also showed inhibitory activity in Vero E6 cells (\cite{weston2020broad}).

Several of the drugs we identified have been reported to potentially act on the SARS-CoV-2 replication: doramapimod (\cite {raymonda2020pharmacologic}), lapatinib (\cite {raymonda2020pharmacologic}), ponatinib (\cite {sauvat2020target}). Imatinib has been shown to be an inhibitor of SARS-CoV-2 entry on a lung organoid model using human pluripotent stem cells (\cite{han2021identification}).

There are also drugs with conflicting results based on the type of cells used, such as bosutinib which is active in Huh 7, active in Vero E6, but not as potent, and inactive in Cacao-2 E6 and in iAEC2 \cite {mirabelli2020morphological}.

BRAF inhibitors (dabrafenib, regorafenib and sorafenib) as well as baricitinib lead to increased virus growth (\cite{stukalov2021multi}). Regorafenib may play a role in the receptor mediated host response to SARS-CoV-2 \cite{ellinger2021sars}, dabrafenib may inhibit SARS-CoV-2 infection \cite{wan2020high} and sorafenib could prevent in vitro replication \cite{klann2020growth}. According to \cite{ellinger2021sars}, sorafenib has antiviral activities, but the cytotoxic and antiviral IC50 values are close. Baricitinib prevented progression to a severe form by modulating the patient's immune landscape in a 20-case study \cite{bronte2020baricitinib} and its administration resulted in an improvement in respiratory function in a study of 62 patients who received baricitinib and coticosteroids compared with 50 patients who received only corticosteroids \cite{rodriguez2021baricitinib}. In another study \cite {hoang2021baricitinib} in rhesus macaques, baricitinib suppressed the production of proinflammatory cytokines, mentained innate antiviral responses and SARS-CoV-2 T cells, and limited the recruitment of neutrophils to the lungs and neutrophil cell death (NETosis).

\paragraph{Clinical trials.} We checked which of the drugs identified in our analyses have been included in clinical trials for COVID-19. We looked for them first in the dedicated COVID-19 section of \cite{drugbank}. We then verified ClinicalTrails.gov \cite{zarin2011clinicaltrials}, but also IRTC \cite{solaymani2009iranian} to identify drugs that weren't listed yet on DrugBank. Of the 130 drugs we identified in our analyses, 23 have been investigated in clinical trials. We  included in the supplementary material, for each of the validated drugs, a selection of clinical trials dedicated to them. Of the 23 drugs included in clinical trials, 14 were obtained only on the MOI 0.01 network, 8 on both networks and one, ibuprofen, only on MOI 0.3 network.

Ibuprofen was obtained due to its effect on BCL2. It is currently included in two clinical trials (NCT04382768, NCT04334629) to be evaluated for its ability to reduce the severity and progression of lung injury.

The followings drugs were obtained only on the MOI 0.01 network: argatroban, baricitinib, bivalirudin, estradiol, estradiol cypionate, imatinib, lidocaine, lithium carbonate, nafamostat, ruxolitinib, suramin, tamoxifen, tofacitinib, zanubrutinib. Argatroban and bivalirudin may help in COVID-19 due to their action on coagulation, especially when heparin resistance is involved \cite{arachchillage2020anticoagulation}, \cite{seelhammer2020covid}. Argatroban is included in NCT04406389 clinical trail together with enoxaparin, fondapariniux and unfractioned heparin to establish if they can reduce mortality, while bivalirudin is included in NCT04445935, to see if it could prevent clotting better than low molecular weight heparin and support the dissolution of existing clots. Nafamostat, another F2 inhibitor, could be used due to its antiviral, anti-inflammatory and anti-coagulation properties (NCT04418128). According to clinical studies, estradiol may control the inflammmation (IRCT20150716023235N15), while lidocaine may impact the gas exchange and inflammation in acute respiratory distress syndrome (NCT04609865). Tofacitinib, a selective immunosuppressant used in rheumatoid arthritis that is also being investigated for its use in preventing transplant rejection \cite{drugbank}, is included in clinical studies to find out if it is suitable to moderate or severe cases (NCT04469114, NCT04415151). Imatinib may have some antiviral properties due to lysosomal alkalization (NCT04394416), while zanubrutinib may lower elevated levels of pro‐inflammatory cytokines \cite{thibaud2020protective}, \cite{nascimento2020sars}, and it is investigated as it could increase the respiratory failure-free survival rate (NCT04382586). Lithium carbonate could also have an effect on pro-inflammatory cytokines \cite{qaswal2021potential}, and it is included in several clinical trials, such as: CTRI/2020/06/026193, IRCT20081019001369N5, IRCT20130812014333N147. Tamoxifen is administred in two clinical studies, but only in combination with other drugs (NCT04389580, NCT04568096).

The drugs obtained on both networks are: acetylsalicylic acid, arsenic trioxide, dasatinib, genistein, minocycline, quercetin, resveratrol and sirolimus. Aspirin has been investigated in COVID-19 clinical trials for its use in coagulation problems (NCT04363840, NCT04498273), but may inhibit virus replication and reduce lung injury (NCT04365309). It inhibits the platelet aggregation due to its action on COX-1. However, we did not identify it through its effect on COX-1, but through its action on TP53, on RPS6KA3, on PRKAA1, and on MAPK1. The only drug target obtained for both MOI network analyses is TP53. The action on TP53 could be related to the aspirin-induced inhibition of adipogenesis \cite{su2014aspirin}; obesity can aggravate COVID-19.

Other clinical trials have mentioned possible effects for the followings drugs: genistein (could be efficient in pulmonary fibrosis in patients discharged from hospital (NCT04482595)), quercetin (may help because it has antiviral activities in SARS infection and antioxidant properties (NCT04377789)), resveratrol (could have anti-fibrotic effects (NCT04799743)),  sirolimus (may help in reducing lung injury/acute respiratory distress syndrome (NCT04482712)), dasatinib (may help reduce the strong inflammation (NCT04830735)).

\paragraph{Computational validation.} More than half of the drugs identified by our analyses were also found through other computational approaches. We discuss them in the supplementary information, Section~\ref{section-supplementary-information-computational-validation}.

\section{Discussion}
\label{section-discussion}

The recent study of \cite{daniloski2021identification} on the survivability of SARS-CoV-2 infected cells identified many host factors that are essential for the SARS-CoV-2 infection. This offers a new guide to therapeutic targeting of COVID-19. Yet, out of their top 200 ranked genes in each of the two viral load experiments only 23 (24, resp.) are drug targetable \cite{drugbank}. We investigated whether they can be instead targeted through short network signaling pathways and found 40 drug targetable proteins that can control these genes. We identified 130 drugs that target these proteins and can be significant in COVID-19. Some of these include drugs recommended to be used, drugs that have been evaluated in clinical trials and various in-vitro assays. The network-based approach offers a wider spectrum of options by identifying drug targetable nodes at a short distance upstream of the SARS-CoV-2 infection host factors, able to influence them in the sense of network controllability. Moreover, we also identified in this way possible mechanisms of action for drugs, offering a mechanistic understandings of drugs through the changes they may induce in the protein-protein interactions.

Many of the drugs we identified act on the immune system (including the FDA-approved COVID-19 drug baricitinib), and on the coagulation cascade (e.g. nafamostat). Antivirals were not found by our analyses because for many of them (e.g. favipiravir, remdesivir, umifenovir) their human targets are unknown.  Our list of drugs also includes false positives, such as drugs that are applied topically and should not be swallowed (e.g. ingenol mebutate), but also possible harmful substances in this disease (kappadione). They are found by our algorithms because they do influence the host factors, in addition to their other effects. Our results could be further improved by leveraging data on the quantitative strength of various interactions, on the result of their concurrent activation/inhibition signals, and on the specificity of their mechanism in the context of the disease. Such data would be needed for all the protein-protein interactions in our network, not just for a selected subset of them. A quantitative version of network controllability theory remains to be developed to scale to become applicable to such data.

The network-based approach explored in this study can be applicable to other diseases and can be especially fruitful in drug repurposing for rare diseases. The key part of our approach is identifying a set of  targets, whose control may be therapeutically beneficial. In this study we used as control targets the host factors required for SARS-CoV-2 infections. The network controllability analysis yields a set of input nodes that can control these targets. Moreover, the input nodes can be selected to a large extent to be drug targetable with currently available drugs, bringing this method within the drug repurposing realm. Each of the input nodes identified by the analysis (and the drugs targeting them), as well as various combinations of them, can be used to influence some of the control targets. This yields a rich set of predictions that could inform the setup of new drug repurposing clinical trials.

\section{Materials and methods}
\label{section-materials-and-methods}

\subsection{Data}
\label{section-materials-and-methods-data}

The signaling data was extracted from the KEGG \cite{kanehisa:2000aa}, OmniPath \cite{turei:2016aa} and SIGNOR \cite{10.1093/nar/gkz949} databases. Only the directed interactions were considered. An interaction can appear in multiple databases. The interactions were matched based on the UniProt identifiers of their proteins, which are provided by default by all of the three databases.

We considered as targets of our study the $200$ highest ranked host factors in the cell survivability experiments of \cite{daniloski2021identification} at 0.01 MOI and 0.3 MOI that were shown to be required for SARS-CoV-2 infection. These proteins were targeted through short multi-step signaling paths originating in drug targets. The drug targets were collected from the DrugBank \cite{drugbank} database. We selected the drug-targets of the approved and investigational small molecule drugs, except for those illicit or nutraceutical. We also discarded a number of specific $19$ approved and/or investigational drugs that have more than $50$ targets, most of which are not targeted by any other drug.

\subsection{Network generation}
\label{section-materials-and-methods-network-generation}

For each of the two MOI experiments of \cite{daniloski2021identification} we identified all proteins that are upstream of the top 200 highest ranked host factors, at a distance of maximum two interactions. We also identified all proteins that are downstream of the drug-target proteins, at a distance of maximum two interactions. The network generation is discussed in all details in the supplementary information, Section~\ref{section-supplementary-information-network}.

\subsection{Network analysis}
\label{section-materials-and-network-analysis}

We used the NetControl4BioMed web application (\cite{netcontrol4biomed2021}) to apply the target controllability analysis on the two generated networks using the greedy algorithm described in \cite{Kanhaiya2017}. The analysis setup is the same in both cases: we use the set of host factors in each network as controllability targets and we aim on identifying a minimal set of proteins that control the host factors, while maximizing the number of drug-target proteins among them. We are limiting the analysis to using control paths of maximum length $3$ (i.e. there are at most three interactions between a drug-target and a host factor that it controls), to minimize the potential dissipation of a drug's effects along the pathway. The analyses are stochastic, which means that the same input can lead to different control solutions.

For each control analysis, we define the control score of a drug-target as the number of host factors it controls in that analysis. We then define the efficiency score of a drug as the maximum control score of any of its drug targets. Thus, for each analysis, we identify the set of drugs with the maximum efficiency score. The actual value of the maximum score may vary between analyses. We then define the efficiency score of a drug as the maximum control score of its set of drug targets within any of the stochastic analyses we ran. Thus, in each analysis, we identify the drugs with the maximum efficiency score. The actual value of the maximum score may vary between analyses.

We run repeatedly the network controllability analysis for each network until no new top ranked drug is found within three consecutive runs.

\section{Data Availability}
\label{section-data-availability}

The drug-target data set used in this article, as well as the generated networks, are available online in the GitHub repository  \cite{inbetweennetgeneration2021}.

\section*{Acknowledgments}

This work was partially supported by the Romanian Ministry of Education and Research, CCCDI – UEFISCDI, project number PNIII-P2-2.1-PED-2019-2391, within PNCDI III.

\bibliographystyle{unsrt}

\bibliography{bibliography}

\begin{thebibliography}{100}

\bibitem{Dong:2020te}
Ensheng Dong et~al.
\newblock An interactive web-based dashboard to track {{COVID-19}} in real
  time.
\newblock {\em The Lancet Infectious Diseases}, 20(5):533--534, 2021/05/12
  2020.

\bibitem{gordon2020sars}
David~E. Gordon et~al.
\newblock A {SARS-CoV-2} protein interaction map reveals targets for drug
  repurposing.
\newblock {\em Nature}, 583(7816):459--468, 2020.

\bibitem{daniloski2021identification}
Zharko Daniloski et~al.
\newblock Identification of required host factors for {SARS-CoV-2} infection in
  human cells.
\newblock {\em Cell}, 184(1):92--105.e16, January 2021.

\bibitem{Tomazou:2021uk}
Marios Tomazou et~al.
\newblock Multi-omics data integration and network-based analysis drives a
  multiplex drug repurposing approach to a shortlist of candidate drugs against
  {{COVID-19}}.
\newblock {\em Briefings in Bioinformatics}, 5/12/2021 2021.

\bibitem{drugbank}
David~S Wishart et~al.
\newblock Drugbank 5.0: a major update to the drugbank database for 2018.
\newblock {\em Nucleic Acids Research}, 46:1074--1082, 2017.

\bibitem{liu:2011}
Yang-Yu Liu et~al.
\newblock Controllability of complex networks.
\newblock {\em Nature}, 2011.

\bibitem{netcontrol4biomed2021}
Victor Popescu et~al.
\newblock {NetControl4BioMed}.
\newblock Available at \url{https://netcontrol.combio.org/}, 2021.

\bibitem{kanehisa:2000aa}
Minoru Kanehisa and Susumu Goto.
\newblock Kegg: kyoto encyclopedia of genes and genomes.
\newblock {\em Nucleic Acids Res}, 28(1):27--30, Jan 2000.

\bibitem{turei:2016aa}
D{\'e}nes T{\"u}rei et~al.
\newblock Omnipath: guidelines and gateway for literature-curated signaling
  pathway resources.
\newblock {\em Nature Methods}, 13(12):966--967, 2016.

\bibitem{10.1093/nar/gkz949}
Luana Licata et~al.
\newblock {SIGNOR 2.0, the SIGnaling Network Open Resource 2.0: 2019 update}.
\newblock {\em Nucleic Acids Research}, 10 2019.
\newblock gkz949.

\bibitem{10.1093/bib/bbz011}
Xiangrong Liu et~al.
\newblock {Computational methods for identifying the critical nodes in
  biological networks}.
\newblock {\em Briefings in Bioinformatics}, 21(2):486--497, 02 2019.

\bibitem{pmid32778962}
E.~Weisberg et~al.
\newblock {{R}epurposing of {K}inase {I}nhibitors for {T}reatment of
  {C}{O}{V}{I}{D}-19}.
\newblock {\em Pharm Res}, 37(9):167, Aug 2020.

\bibitem{Ma-LauerE5192}
Yue Ma-Lauer et~al.
\newblock p53 down-regulates {SARS} coronavirus replication and is targeted by
  the {SARS}-unique domain and {PLpro} via {E3} ubiquitin ligase {RCHY1}.
\newblock {\em Proceedings of the National Academy of Sciences},
  113(35):E5192--E5201, 2016.

\bibitem{Appelberg2020}
Sofia Appelberg et~al.
\newblock Dysregulation in {Akt/mTOR/HIF-1} signaling identified by
  proteo-transcriptomics of {{SARS-CoV-2}} infected cells.
\newblock {\em Emerging microbes {\&} infections}, 9(1):1748--1760, Dec 2020.
\newblock PMC7473213[pmcid].

\bibitem{Gargouri2021}
Mohamed Gargouri et~al.
\newblock Cyclin dependent kinase inhibitors as a new potential therapeutic
  option in management of {COVID-19}.
\newblock {\em Medical hypotheses}, 146:110380--110380, Jan 2021.
\newblock PMC7649033[pmcid].

\bibitem{Dolan2020}
Mary~E. Dolan et~al.
\newblock Investigation of {COVID-19} comorbidities reveals genes and pathways
  coincident with the {SARS-CoV-2} viral disease.
\newblock {\em Scientific Reports}, 10(1):20848, Nov 2020.

\bibitem{uniprot2015uniprot}
The~UniProt Consortium.
\newblock {UniProt: the universal protein knowledgebase in 2021}.
\newblock {\em Nucleic Acids Research}, 49(D1):D480--D489, 11 2020.

\bibitem{Codo2020}
Ana~Campos Codo et~al.
\newblock Elevated glucose levels favor {SARS-CoV-2} infection and monocyte
  response through a {HIF}-1$\alpha$/glycolysis-dependent axis.
\newblock {\em Cell Metabolism}, 32(3):437--446.e5, September 2020.

\bibitem{klann2020growth}
Kevin Klann et~al.
\newblock Growth factor receptor signaling inhibition prevents {SARS-CoV-2}
  replication.
\newblock {\em Molecular Cell}, 80(1):164--174.e4, 2020.

\bibitem{allegra2020immunopathology}
Alessandro Allegra et~al.
\newblock Immunopathology of {{SARS-CoV-2}} infection: Immune cells and
  mediators, prognostic factors, and immune-therapeutic implications.
\newblock {\em International Journal of Molecular Sciences}, 21(13), 2020.

\bibitem{romanelli2020sirolimus}
Antonio Romanelli and Silvia Mascolo.
\newblock Sirolimus to treat {SARS-CoV-2} infection: an old drug for a new
  disease.
\newblock {\em J Res Clin Med}, 8(1):44--44, 2020.

\bibitem{giannis2020coagulation}
Dimitrios Giannis et~al.
\newblock Coagulation disorders in coronavirus infected patients: {COVID-19},
  {SARS-CoV-1}, {MERS-CoV} and lessons from the past.
\newblock {\em Journal of Clinical Virology}, 127:104362, 2020.

\bibitem{costello2019disseminated}
Ryan~A. Costello and Sara~M. Nehring.
\newblock Disseminated intravascular coagulation., Jan 2021.

\bibitem{connors2020thromboinflammation}
Jean~M. Connors and Jerrold~H. Levy.
\newblock Thromboinflammation and the hypercoagulability of {{COVID-19}}.
\newblock {\em Journal of Thrombosis and Haemostasis}, 18(7):1559--1561, 2020.

\bibitem{janssen2020vitamin}
Rob Janssen et~al.
\newblock Vitamin {K} metabolism as the potential missing link between lung
  damage and thromboembolism in coronavirus disease 2019.
\newblock {\em British Journal of Nutrition}, pages 1--8, 2020.

\bibitem{yang2020clinical}
Xiaobo Yang et~al.
\newblock Clinical course and outcomes of critically ill patients with
  {SARS-CoV-2} pneumonia in {Wuhan}, {China}: a single-centered, retrospective,
  observational study.
\newblock {\em The Lancet Respiratory Medicine}, 8(5):475--481, May 2020.

\bibitem{kovats2015estrogen}
Susan Kovats.
\newblock Estrogen receptors regulate innate immune cells and signaling
  pathways.
\newblock {\em Cellular Immunology}, 294(2):63--69, 2015.
\newblock Sex Disparity in Immune Responses.

\bibitem{la2020sex}
Sandro La~Vignera et~al.
\newblock Sex-specific {SARS-CoV-2} mortality: Among hormone-modulated {ACE2}
  expression, risk of venous thromboembolism and hypovitaminosis {D}.
\newblock {\em International Journal of Molecular Sciences}, 21(8), 2020.

\bibitem{seeland2020evidence}
Ute Seeland et~al.
\newblock Evidence for treatment with estradiol for women with {SARS-CoV-2}
  infection.
\newblock {\em BMC Medicine}, 18(1):369, 2020.

\bibitem{middleton1998effect}
Elliot~Jr Middleton and Chithan Kandaswami.
\newblock Effects of flavonoids on immune and inflammatory cell functions.
\newblock {\em Biochemical pharmacology}, 43:1167--79, Mar 1992.

\bibitem{nijveldt2001flavonoids}
Robert~J Nijveldt et~al.
\newblock {Flavonoids: a review of probable mechanisms of action and potential
  applications}.
\newblock {\em The American Journal of Clinical Nutrition}, 74(4):418--425, 10
  2001.

\bibitem{bouhaddou2020global}
Mehdi Bouhaddou et~al.
\newblock The global phosphorylation landscape of {SARS-CoV-2} infection.
\newblock {\em Cell}, 182(3):685--712.e19, August 2020.

\bibitem{gilroy2020treating}
Derek~W. Gilroy et~al.
\newblock Treating exuberant, non-resolving inflammation in the lung;
  implications for acute respiratory distress syndrome and {COVID-19}.
\newblock {\em Pharmacology \& Therapeutics}, page 107745, 2020.

\bibitem{Sachs2019}
Julia Sachs et~al.
\newblock Novel 3,4-dihydroisocoumarins inhibit human {P-gp} and {BCRP} in
  multidrug resistant tumors and demonstrate substrate inhibition of {Yeast}
  {Pdr5}.
\newblock {\em Frontiers in pharmacology}, 10(31040786):400--400, April 2019.

\bibitem{Saccon2021}
Tatiana~Dandolini Saccon et~al.
\newblock Senolytic combination of dasatinib and quercetin alleviates
  intestinal senescence and inflammation and modulates the gut microbiome in
  aged mice.
\newblock {\em The journals of gerontology. Series A, Biological sciences and
  medical sciences}, Jan 2021.

\bibitem{Iacovelli2015}
Stefano Iacovelli et~al.
\newblock Co-targeting of {Bcl-2} and {mTOR} pathway triggers synergistic
  apoptosis in {BH3} mimetics resistant acute lymphoblastic leukemia.
\newblock {\em Oncotarget}, 6(26392332):32089--32103, October 2015.

\bibitem{brimacombe2020opendata}
Kyle~R. Brimacombe et~al.
\newblock An {OpenData portal} to share {{COVID-19}} drug repurposing data in
  real time.
\newblock {\em bioRxiv}, 2020.

\bibitem{giudice2020combination}
Valentina Giudice et~al.
\newblock Combination of ruxolitinib and eculizumab for treatment of severe
  {SARS-CoV-2}-related acute respiratory distress syndrome: A controlled study.
\newblock {\em Frontiers in Pharmacology}, 11:857, 2020.

\bibitem{novartis_2020}
Novartis.
\newblock Novartis provides update on {RUXCOVID} study of ruxolitinib for
  hospitalized patients with {COVID-19}, Dec 2020.

\bibitem{abedi2020plausibility}
Farshad Abedi et~al.
\newblock Plausibility of therapeutic effects of {Rho} kinase inhibitors
  against {Severe} {Acute} {Respiratory} {Syndrome} {Coronavirus} 2
  ({{COVID-19}}).
\newblock {\em Pharmacological Research}, 156:104808, 2020.

\bibitem{ghafouri2020angiotensin}
Soudeh Ghafouri-Fard et~al.
\newblock Angiotensin converting enzyme: A review on expression profile and its
  association with human disorders with special focus on {SARS-CoV-2}
  infection.
\newblock {\em Vascular Pharmacology}, 130:106680, 2020.

\bibitem{gouda2020snake}
Ahmed~S. Gouda and Bruno M{\'e}garbane.
\newblock Snake venom-derived bradykinin-potentiating peptides: A promising
  therapy for {COVID-19}?
\newblock {\em Drug Development Research}, 82(1):38--48, 2021.

\bibitem{hoffmann2020nafamostat}
Markus Hoffmann et~al.
\newblock Nafamostat mesylate blocks activation of {SARS-CoV-2}: New treatment
  option for {COVID-19}.
\newblock {\em Antimicrobial Agents and Chemotherapy}, 64(6), 2020.

\bibitem{yamamoto2020anticoagulant}
Mizuki Yamamoto et~al.
\newblock The anticoagulant nafamostat potently inhibits {SARS-CoV-2} {S}
  protein-mediated fusion in a cell fusion assay system and viral infection in
  vitro in a cell-type-dependent manner.
\newblock {\em Viruses}, 12(6), 2020.

\bibitem{ko2021comparative}
Meehyun Ko et~al.
\newblock Comparative analysis of antiviral efficacy of {FDA}-approved drugs
  against {SARS-CoV-2} in human lung cells.
\newblock {\em Journal of Medical Virology}, 93(3):1403--1408, 2021.

\bibitem{doi2020nafamostat}
Kent Doi et~al.
\newblock Nafamostat mesylate treatment in combination with favipiravir for
  patients critically ill with {Covid-19}: a case series.
\newblock {\em Critical Care}, 24(1):392, 2020.

\bibitem{cagno2020tyrosine}
Valeria Cagno et~al.
\newblock The tyrosine kinase inhibitor nilotinib inhibits {SARS-CoV-2} in
  vitro.
\newblock {\em Basic \& Clinical Pharmacology \& Toxicology}, 128(4):621--624,
  2021.

\bibitem{varghese2021berberine}
Finny~S. Varghese et~al.
\newblock Berberine and obatoclax inhibit {SARS-Cov-2} replication in primary
  human nasal epithelial cells in vitro.
\newblock {\em Viruses}, 13(2), 2021.

\bibitem{ianevski2020potential}
Aleksandr Ianevski et~al.
\newblock Potential antiviral options against {SARS-CoV-2} infection.
\newblock {\em Viruses}, 12(6), 2020.

\bibitem{weston2020broad}
Stuart Weston et~al.
\newblock Broad anti-coronavirus activity of food and drug
  administration-approved drugs against {SARS-CoV-2} in vitro and {SARS-CoV} in
  vivo.
\newblock {\em Journal of Virology}, 94(21), 2020.

\bibitem{raymonda2020pharmacologic}
Matthew~H. Raymonda et~al.
\newblock Pharmacologic profiling reveals lapatinib as a novel antiviral
  against {SARS-CoV-2} in vitro.
\newblock {\em bioRxiv}, 2020.

\bibitem{sauvat2020target}
Allan Sauvat et~al.
\newblock On-target versus off-target effects of drugs inhibiting the
  replication of {SARS-CoV-2}.
\newblock {\em Cell Death \& Disease}, 11(8):656, 2020.

\bibitem{han2021identification}
Yuling Han et~al.
\newblock Identification of {SARS-CoV-2} inhibitors using lung and colonic
  organoids.
\newblock {\em Nature}, 589(7841):270--275, 2021.

\bibitem{mirabelli2020morphological}
Carmen Mirabelli et~al.
\newblock Morphological cell profiling of {SARS-CoV-2} infection identifies
  drug repurposing candidates for {COVID-19}.
\newblock {\em bioRxiv}, 2020.

\bibitem{stukalov2021multi}
Alexey Stukalov et~al.
\newblock Multilevel proteomics reveals host perturbations by {SARS-CoV-2} and
  {SARS-CoV}.
\newblock {\em Nature}, 2021.

\bibitem{ellinger2021sars}
Bernhard Ellinger et~al.
\newblock A {SARS-CoV-2} cytopathicity dataset generated by high-content
  screening of a large drug repurposing collection.
\newblock {\em Scientific Data}, 8(1):70, 2021.

\bibitem{wan2020high}
Weiwei Wan et~al.
\newblock High-throughput screening of an {FDA}-approved drug library
  identifies inhibitors against arenaviruses and {SARS-CoV-2}.
\newblock {\em ACS Infect. Dis.}, November 2020.

\bibitem{bronte2020baricitinib}
Vincenzo Bronte et~al.
\newblock Baricitinib restrains the immune dysregulation in patients with
  severe {{COVID-19}}.
\newblock {\em The Journal of Clinical Investigation}, 130(12):6409--6416, 12
  2020.

\bibitem{rodriguez2021baricitinib}
Jose~Luis Rodriguez-Garcia et~al.
\newblock {Baricitinib improves respiratory function in patients treated with
  corticosteroids for {SARS-CoV-2} pneumonia: an observational cohort study}.
\newblock {\em Rheumatology}, 60(1):399--407, 10 2020.

\bibitem{hoang2021baricitinib}
Timothy~N. Hoang et~al.
\newblock Baricitinib treatment resolves lower-airway macrophage inflammation
  and neutrophil recruitment in {SARS-CoV-2}-infected rhesus macaques.
\newblock {\em Cell}, 184(2):460--475.e21, January 2021.

\bibitem{zarin2011clinicaltrials}
Deborah~A. Zarin et~al.
\newblock The {ClinicalTrials.gov} results database--update and key issues.
\newblock {\em The New England journal of medicine}, 364(21366476):852--860,
  March 2011.

\bibitem{solaymani2009iranian}
Masoud Solaymani-Dodaran et~al.
\newblock Iranian registry of clinical trials: path and challenges from
  conception to a world health organization primary register.
\newblock {\em Journal of Evidence-Based Medicine}, 2(1):32--35, 2009.

\bibitem{arachchillage2020anticoagulation}
Deepa~J. Arachchillage et~al.
\newblock Anticoagulation with argatroban in patients with acute antithrombin
  deficiency in severe {{COVID-19}}.
\newblock {\em British Journal of Haematology}, 190(5):e286--e288, 2020.

\bibitem{seelhammer2020covid}
Troy~G. Seelhammer et~al.
\newblock {COVID-19} and {ECMO}: An unhappy marriage of endothelial dysfunction
  and hemostatic derangements.
\newblock {\em Journal of Cardiothoracic and Vascular Anesthesia},
  34(12):3193--3196, December 2020.

\bibitem{thibaud2020protective}
Santiago Thibaud et~al.
\newblock Protective role of {Bruton} tyrosine kinase inhibitors in patients
  with chronic lymphocytic leukaemia and {COVID-19}.
\newblock {\em British Journal of Haematology}, 190(2):e73--e76, 2020.

\bibitem{nascimento2020sars}
Jos{\'e} Ad{\~a}o Carvalho~Nascimento Junior et~al.
\newblock {SARS}, {MERS} and {SARS-CoV-2} ({COVID-19}) treatment: a patent
  review.
\newblock {\em Expert Opinion on Therapeutic Patents}, 30(8):567--579, 2020.
\newblock PMID: 32429703.

\bibitem{qaswal2021potential}
Abdallah~Barjas Qaswal et~al.
\newblock The potential role of lithium as an antiviral agent against
  {SARS-CoV-2} via membrane depolarization: Review and hypothesis.
\newblock {\em Scientia Pharmaceutica}, 89(1), 2021.

\bibitem{su2014aspirin}
Ying-Fang Su et~al.
\newblock Aspirin-induced inhibition of adipogenesis was p53-dependent and
  associated with inactivation of pentose phosphate pathway.
\newblock {\em European Journal of Pharmacology}, 738:101--110, 2014.

\bibitem{Kanhaiya2017}
Krishna Kanhaiya et~al.
\newblock Controlling directed protein interaction networks in cancer.
\newblock {\em Scientific Reports}, 7, 2017.

\bibitem{inbetweennetgeneration2021}
Victor Popescu.
\newblock {PPI} network generation code.
\newblock Available at \url{https://github.com/Vilksar/InBetweenNetGeneration},
  2021.

\bibitem{Liu2011}
Yang-Yu Liu et~al.
\newblock Controllability of complex networks.
\newblock {\em Nature}, 2011.

\bibitem{Gao:2014aa}
Jianxi Gao et~al.
\newblock Target control of complex networks.
\newblock {\em Nature Communications}, 5(1):5415, 2014.

\bibitem{cze18}
Eugen Czeizler et~al.
\newblock Structural target controllability of linear networks.
\newblock {\em IEEE/ACM Transactions on Computational Biology and
  Bioinformatics}, 15:1217--1228, 2018.

\bibitem{Kanhaiya2018}
Krishna Kanhaiya et~al.
\newblock {NetControl4BioMed}: a pipeline for biomedical data acquisition and
  analysis of network controllability.
\newblock {\em BMC Bioinformatics}, 19(7):185, Jul 2018.

\bibitem{Guo:2017aa}
Wei-Feng Guo et~al.
\newblock Constrained target controllability of complex networks.
\newblock {\em Journal of Statistical Mechanics: Theory and Experiment},
  2017(6):063402, 2017.

\bibitem{eleftheriou2020silico}
Phaedra Eleftheriou et~al.
\newblock In silico evaluation of the effectivity of approved protease
  inhibitors against the main protease of the novel {SARS-CoV-2} virus.
\newblock {\em Molecules}, 25(11), 2020.

\bibitem{chakraborti2020repurposing}
Sohini Chakraborti et~al.
\newblock Repurposing drugs against the main protease of {SARS-CoV-2}:
  mechanism-based insights supported by available laboratory and clinical data.
\newblock {\em Mol. Omics}, 16:474--491, 2020.

\bibitem{qiao2020computational}
Zhen Qiao et~al.
\newblock Computational view toward the inhibition of {SARS-CoV-2} spike
  glycoprotein and the {3CL} protease.
\newblock {\em Computation}, 8(2), 2020.

\bibitem{pandey2020silico}
Anand~Kumar Pandey and Shalja Verma.
\newblock An in-silico evaluation of dietary components for structural
  inhibition of {SARS-Cov-2} main protease.
\newblock {\em Journal of Biomolecular Structure and Dynamics}, 0(0):1--7,
  2020.
\newblock PMID: 32811367.

\bibitem{falade2021silico}
Victoria~Adeola Falade et~al.
\newblock In silico investigation of saponins and tannins as potential
  inhibitors of {SARS-CoV-2} main protease {(Mpro)}.
\newblock {\em In silico pharmacology}, 9:9, 2021.

\bibitem{olubiyi2020high}
Olujide~O. Olubiyi et~al.
\newblock High throughput virtual screening to discover inhibitors of the main
  protease of the coronavirus {SARS-CoV-2}.
\newblock {\em Molecules}, 25(14), 2020.

\bibitem{elzupir2020caffeine}
Amin~O. Elzupir.
\newblock Caffeine and caffeine-containing pharmaceuticals as promising
  inhibitors for 3-chymotrypsin-like protease of {SARS-CoV-2}.
\newblock {\em Journal of Biomolecular Structure and Dynamics}, 0(0):1--8,
  2020.
\newblock PMID: 33094705.

\bibitem{kandeel2020repurposing}
Mahmoud Kandeel et~al.
\newblock Repurposing of {FDA}-approved antivirals, antibiotics, anthelmintics,
  antioxidants, and cell protectives against {SARS-CoV-2} papain-like protease.
\newblock {\em Journal of Biomolecular Structure and Dynamics}, 0(0):1--8,
  2020.
\newblock PMID: 32597315.

\bibitem{kouznetsova2020potential}
Valentina~L. Kouznetsova et~al.
\newblock Potential {COVID-19} papain-like protease {PLpro} inhibitors:
  repurposing {FDA}-approved drugs.
\newblock {\em PeerJ}, 8(32999768):e9965--e9965, September 2020.

\bibitem{machitani2020rna}
Mitsuhiro Machitani et~al.
\newblock {RNA}-dependent {RNA} polymerase, {RdRP}, a promising therapeutic
  target for cancer and potentially {COVID-19}.
\newblock {\em Cancer Science}, 111(11):3976--3984, 2020.

\bibitem{ruan2021sars}
Zijing Ruan et~al.
\newblock {SARS-CoV-2} and sars-cov: Virtual screening of potential inhibitors
  targeting rna-dependent rna polymerase activity (nsp12).
\newblock {\em Journal of Medical Virology}, 93(1):389--400, 2021.

\bibitem{alexpandi2020quinolines}
Rajaiah Alexpandi et~al.
\newblock Quinolines-based {{SARS-CoV-2}} {3CLpro} and {RdRp} inhibitors and
  {Spike-RBD-ACE2} inhibitor for drug-repurposing against {{COVID-19}}: An in
  silico analysis.
\newblock {\em Frontiers in Microbiology}, 11:1796, 2020.

\bibitem{buitron2021silico}
Ivonne Buitr{\'o}n-Gonz{\'a}lez et~al.
\newblock In-silico drug repurposing study: Amprenavir, enalaprilat, and
  plerixafor, potential drugs for destabilizing the {SARS-CoV-2}
  s-protein-angiotensin-converting enzyme 2 complex.
\newblock {\em Results in Chemistry}, 3:100094, 2021.

\bibitem{trezza2020integrated}
Alfonso Trezza et~al.
\newblock An integrated drug repurposing strategy for the rapid identification
  of potential {SARS-CoV-2} viral inhibitors.
\newblock {\em Scientific Reports}, 10(1):13866, 2020.

\bibitem{kalathiya2020highly}
Umesh Kalathiya et~al.
\newblock Highly conserved homotrimer cavity formed by the {SARS-CoV-2} spike
  glycoprotein: A novel binding site.
\newblock {\em Journal of Clinical Medicine}, 9(5), 2020.

\bibitem{maffucci2020silico}
Irene Maffucci and Alessandro Contini.
\newblock In silico drug repurposing for {SARS-CoV-2} main proteinase and spike
  proteins.
\newblock {\em J. Proteome Res.}, 19(11):4637--4648, November 2020.

\bibitem{awad2020high}
Ibrahim~E. Awad et~al.
\newblock High-throughput virtual screening of drug databanks for potential
  inhibitors of {{SARS-CoV-2}} spike glycoprotein.
\newblock {\em Journal of Biomolecular Structure and Dynamics}, 0(0):1--14,
  2020.
\newblock PMID: 33103586.

\bibitem{white2020discovery}
Mark~Andrew White et~al.
\newblock Discovery of {COVID-19} inhibitors targeting the {SARS-CoV-2} nsp13
  helicase.
\newblock {\em J. Phys. Chem. Lett.}, 11(21):9144--9151, November 2020.

\bibitem{culletta2020exploring}
Giulia Culletta et~al.
\newblock Exploring the {SARS-CoV-2} proteome in the search of potential
  inhibitors via structure-based pharmacophore modeling/docking approach.
\newblock {\em Computation}, 8(3), 2020.

\bibitem{narayanan2021ritonavir}
Naveen Narayanan and Deepak~T. Nair.
\newblock Ritonavir may inhibit exoribonuclease activity of nsp14 from the
  {SARS-CoV-2} virus and potentiate the activity of chain terminating drugs.
\newblock {\em International Journal of Biological Macromolecules},
  168:272--278, 2021.

\bibitem{al2020prediction}
Nihad A.~M. Al-Rashedi et~al.
\newblock Prediction of potential inhibitors against {{SARS-CoV-2}}
  endoribonuclease: {RNA} immunity sensing.
\newblock {\em Journal of Biomolecular Structure and Dynamics}, 0(0):1--14,
  2020.
\newblock PMID: 33357040.

\bibitem{jiang2020structural}
Yuanyuan Jiang et~al.
\newblock Structural analysis, virtual screening and molecular simulation to
  identify potential inhibitors targeting 2'-o-ribose methyltransferase of
  {SARS-CoV-2} coronavirus.
\newblock {\em Journal of Biomolecular Structure and Dynamics}, 0(0):1--16,
  2020.
\newblock PMID: 33016237.

\bibitem{sadegh2020exploring}
Sepideh Sadegh et~al.
\newblock Exploring the {SARS-CoV-2} virus-host-drug interactome for drug
  repurposing.
\newblock {\em Nature Communications}, 11(1):3518, 2020.

\bibitem{verstraete2020covmulnet19}
Nina Verstraete et~al.
\newblock Covmulnet19, integrating proteins, diseases, drugs, and symptoms: A
  network medicine approach to {COVID-19}.
\newblock {\em Network and Systems Medicine}, 3(1):130--141, 2020.

\bibitem{mousavi2020connectivity}
Seyedeh~Zahra Mousavi et~al.
\newblock A connectivity map-based drug repurposing study and integrative
  analysis of transcriptomic profiling of {SARS-CoV-2} infection.
\newblock {\em Infection, Genetics and Evolution}, 86:104610, 2020.

\bibitem{barh2020multi}
Debmalya Barh et~al.
\newblock Multi-omics-based identification of {SARS-CoV-2} infection biology
  and candidate drugs against {{COVID-19}}.
\newblock {\em Computers in Biology and Medicine}, 126:104051, 2020.

\bibitem{pavel2021integrated}
Alisa Pavel et~al.
\newblock {Integrated network analysis reveals new genes suggesting {COVID-19}
  chronic effects and treatment}.
\newblock {\em Briefings in Bioinformatics}, 22(2):1430--1441, 02 2021.

\bibitem{shetty2020therapeutic}
Rohit Shetty et~al.
\newblock Therapeutic opportunities to manage {COVID-19}/{SARS-CoV-2}
  infection: Present and future.
\newblock {\em Indian Journal of Ophthalmology}, 68(5), 2020.

\bibitem{mckee2020candidate}
Dwight~L. McKee et~al.
\newblock Candidate drugs against {SARS-CoV-2} and {COVID-19}.
\newblock {\em Pharmacological Research}, 157:104859, 2020.

\bibitem{khodadadi2020study}
Ehsaneh Khodadadi et~al.
\newblock Study of combining virtual screening and antiviral treatments of the
  sars-cov-2 (covid-19).
\newblock {\em Microbial Pathogenesis}, 146:104241, 2020.

\bibitem{catalano2021diarylureas}
Alessia Catalano et~al.
\newblock Diarylureas: Repositioning from antitumor to antimicrobials or
  multi-target agents against new pandemics.
\newblock {\em Antibiotics}, 10(1), 2021.

\bibitem{clarke2021appyters}
Daniel~JB Clarke et~al.
\newblock Appyters: Turning jupyter notebooks into data-driven web apps.
\newblock {\em Patterns}, 2(3):100213, 2021.

\bibitem{kuleshov2016enrichr}
Maxim~V Kuleshov et~al.
\newblock Enrichr: a comprehensive gene set enrichment analysis web server 2016
  update.
\newblock {\em Nucleic acids research}, 44(W1):W90--W97, 2016.

\end{thebibliography}

\clearpage

\section{Supplementary information}
\label{section-supplementary-information}

\subsection{The COVID-19 directed protein-protein interaction networks}
\label{section-supplementary-information-network}

\begin{figure}[htb]
    \begin{center}
         \includegraphics[width=\textwidth]{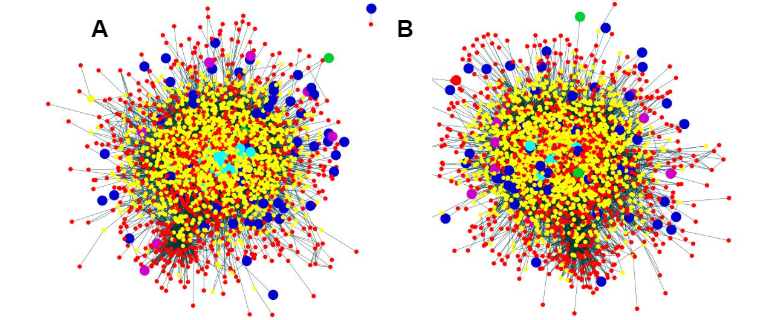}
    \end{center}
    \caption{\textbf{Networks}: \textbf{A} -- MOI 0.01, \textbf{B} -- MOI 0.3. \textbf{Color code}: light blue -- control proteins, green -- control and essential proteins, dark blue -- host factors required for SARS-CoV-2 infection, red -- drug-target proteins, purple --  drug-target proteins that are also host factors, yellow -- remaining proteins}
    \label{figure-networks}
\end{figure}

We constructed two directed protein-protein interaction (PPI) networks, one for each of the datasets of \cite{daniloski2021identification} with the essential host factors required for the SARS-CoV-2 infection at MOI 0.01 and MOI 0.3, resp. We narrowed the interaction networks to include only the proteins upstream of the essential proteins of \cite{daniloski2021identification} and downstream of the drug-targets from DrugBank \cite{drugbank}, at a distance of at most two interactions from either of these sets. We added between these proteins all interactions we found in KEGG, OmniPath and SIGNOR. Any essential protein or drug target that remained isolated in the network was removed. This way, we narrowed the search for possible influences from drug-targets to essential proteins to those acting along very short distances in the networks. The networks are illustrated in Figure \ref{figure-networks} and their topological properties are in Table \ref{table-networks-statistics}. The results of the centrality analysese of the networks are in Table \ref{table-networks-centrality}.

We then calculated the intersection of the two sets of protein identifiers, obtaining all proteins that are on a signaling path from a drug target to an essential gene, at a downstream distance from the drug targets at most 2 and at an upstream distance from the essential genes at most 2. We obtained the networks by considering all the directed interactions between these proteins, the drug target proteins, and the essential proteins. The same method was applied to both MOI datasets, the difference being in the sets of host factors. This obviously also led to differences in the intermediate proteins from the drug targets to the host factors. 

\begin{table}[htb]
    \caption{Statistics of the two protein-protein directed interaction networks. \textbf{N}: number of nodes; \textbf{E}: number of edges; \textbf{T}: number of host factors in the network; \textbf{DT}: number of drug-targets; \textbf{AD}: network average degree; \textbf{D}: network diameter; \textbf{CC}: number of connected components.}
    \label{table-networks-statistics}
    \begin{center}
        \begin{tabular}{c|c|c|c|c|c|c|c}
            \textbf{MOI} & \textbf{N} & \textbf{E} & \textbf{T} & \textbf{DT} & \textbf{AD} & \textbf{D} & \textbf{CC} \\
            \hline
            0.01 & 2491 & 23746 & 70 & 1010 & 17.305 & 10 & 2 \\
            0.3 & 2532 & 24105 & 62 & 1008 & 17.296 & 10 & 1 \\
        \end{tabular}
    \end{center}
\end{table}

\begin{table}[htb!]
    \caption{Top ranked proteins in our networks, based on their centrality measures}
    \label{table-networks-centrality}
    \begin{center}
        \begin{tabular}{c|c|c|c|c|c}
            \textbf{Total degree} & \textbf{In-degree} & \textbf{Out-degree} & \textbf{Closeness} & \textbf{Betweenness} & \textbf{Harmonic}\\
            \hline
			SRC & TP53 & PRKCA & TP53 & SRC & TP53 \\
			GSK3B & PTK2 & SRC & CTNNB1 & TP53 & CTNNB1  \\
			PRKCA & CTNNB1 & GSK3B/CDK1 & SRC & AKT1 & SRC\\
			\hline
        \end{tabular}
    \end{center}
\end{table}

\subsection{Target network controllability}
\label{section-supplementary-information-netcontrol}

Network controllability is about being able to induce arbitrary changes to the variables of a network through suitable changes applied to a (small) set of input nodes. It is a method applicable to linear networks. Such a network with $n\geq 2$ nodes can be represented as an $n$-node directed (weighted) graph, and also as a matrix $A\in \mathbb{R}^{n\times n}$ describing the weights of each possible interaction between the nodes. A directed network is said to be controllable from a set of input nodes if for any desired initial and final numerical configurations of the nodes in the network, there exists a suitable set of input functions applicable to the input nodes that induce the desired change of configuration~\cite{Liu2011}. The methodology accommodates also the possibility of imposing such a desired change of configuration to a target subset of nodes in the network~\cite{Gao:2014aa}. The target network controllability problem asks to find a minimum set of input nodes controlling the target subset.

Formally, we have a dynamical system defined by the system of differential equations $dx(t)/{dt}=Ax(t)$, where $x:\mathbb{R}\rightarrow\mathbb{R}^n$ and $x(t)$ represents the configuration of the $n$ nodes at time $t$. Let $V$ define the $n$-set of nodes of the network. The set of input functions acting on some size-$m$ subset $I\subseteq V$ of the network can be seen as an $m$-dimensional input vector $u$ of real functions, $u:\mathbb{R}\rightarrow\mathbb{R}^m$. Thus, the dynamical system under the influence of these functions is formally described by the system

\begin{equation}
    \label{eq-control}
    dx(t)/{dt}=Ax(t)+B_Iu(t),
\end{equation}

where $B_I\in\mathbb{R}^{n\times m}$ is the characteristic matrix associated to set $I$.

Within this theoretical setting, the target controllability problem for a dynamical system defined by matrix $A\in \mathbb{R}^{n\times n}$ and target set $T\subseteq V$, $|T|=l$, is defined as follows: find the smallest $m\leq n$ and input set (a.k.a controller) $I\subseteq V$, $|I|=m$, such that for any $x(0)\in\mathbb{R}^n$ and any $\alpha\in\mathbb{R}^l$, there is an input vector $u:\mathbb{R}\rightarrow\mathbb{R}^m$ such that the solution $\tilde{x}$ of (\ref{eq-control}) eventually coincides with $\alpha$ on its $T$-components, i.e., $C_T\tilde{x}(\tau)=\alpha$, for some $\tau\geq 0$, where $C_T$ is the characteristic matrix of the target set $T$. Although the problem is known to be computationally hard~\cite{cze18}, some efficient approximation algorithms were provided in~\cite{cze18, Kanhaiya2018}.

Motivated by bio-medical applications of network controllability, a new layer of optimization has been added to the above formulation. Namely, the input-constrained target controllability problem~\cite{Guo:2017aa, Kanhaiya2017}  asks to minimize the size of the controller for a given target $T$ within a network, by maximizing in the same time the use of so-called preferred input nodes within the controller, i.e., elements of a previously defined subset $P\subseteq V$ of preferred input nodes. Indeed such a subset $P$ can be associated in the bio-medical setting to those proteins/genes known to be the targets of available drugs. Thus, it is useful not only to minimize the controller of a targeting subset of focused nodes, but in doing so to rely as much as possible to the use of already available drugs. It is this particular setting of the controllability problem that has been used as the underlying methodological approach in this research.

\subsection{Computational validation}
\label{section-supplementary-information-computational-validation}

More than half of the drugs we found were also reported by other computational studies. Some of these drugs could have an inhibitory effect on the main protease (e.g.	dabigatran etexilate \cite{eleftheriou2020silico}, \cite{chakraborti2020repurposing}, dasatinib \cite{qiao2020computational}, ellagic acid \cite{pandey2020silico}, \cite{falade2021silico}, radotinib \cite{olubiyi2020high}), on the papain-like protease (caffeine \cite{elzupir2020caffeine}, phenformin \cite{kandeel2020repurposing}, ximelagatran \cite{kouznetsova2020potential}), or on the RNA-dependent RNA polymerase (e.g. docetaxel \cite{kandeel2020repurposing}, eribulin \cite{machitani2020rna}, nilotinib \cite{ruan2021sars}, pelitinib \cite{alexpandi2020quinolines}), as well as on the interaction between the spike and ACE2 (e.g. dexibuprofen \cite{buitron2021silico}, midostaurin \cite {trezza2020integrated}, paclitaxel \cite{kalathiya2020highly}, ponatinib \cite{maffucci2020silico}, regorafenib \cite {awad2020high}), or on another viral important point (e.g. nilotinib on nsp13 \cite{white2020discovery}, isoprenaline on nsp9 \cite{culletta2020exploring}, docetaxel on nsp14 \cite{narayanan2021ritonavir}, enzastaurin on nsp15 \cite{al2020prediction}, entrectinib on nsp16 \cite {jiang2020structural}). There are also drugs that can act on host and virus interactions (e.g. erlotinib \cite{sadegh2020exploring}, XL019 \cite{verstraete2020covmulnet19}) or only on host genes (e.g. lidocaine \cite{mousavi2020connectivity}, kappadione \cite{barh2020multi}, \cite{pavel2021integrated}, phenethyl isothiocyanate \cite{barh2020multi}). As we can see, a drug can act on several points. For some of the drugs in our results we found some proposed theories on their potential mode of action (e.g.  bryostatin \cite {nascimento2020sars}, emodin \cite{shetty2020therapeutic}, \cite{mckee2020candidate}, \cite{khodadadi2020study}  or ripretinib \cite {catalano2021diarylureas}). Others can be deduced based on their targets, identified in other studies  (e.g. flovagatran, rimiducid, rindopepimut).

\subsection{Supplementary figures}
\label{section-supplementary-information-supplementary-figures}

\begin{figure}[htb]
    \begin{center}
        \includegraphics[width=\textwidth]{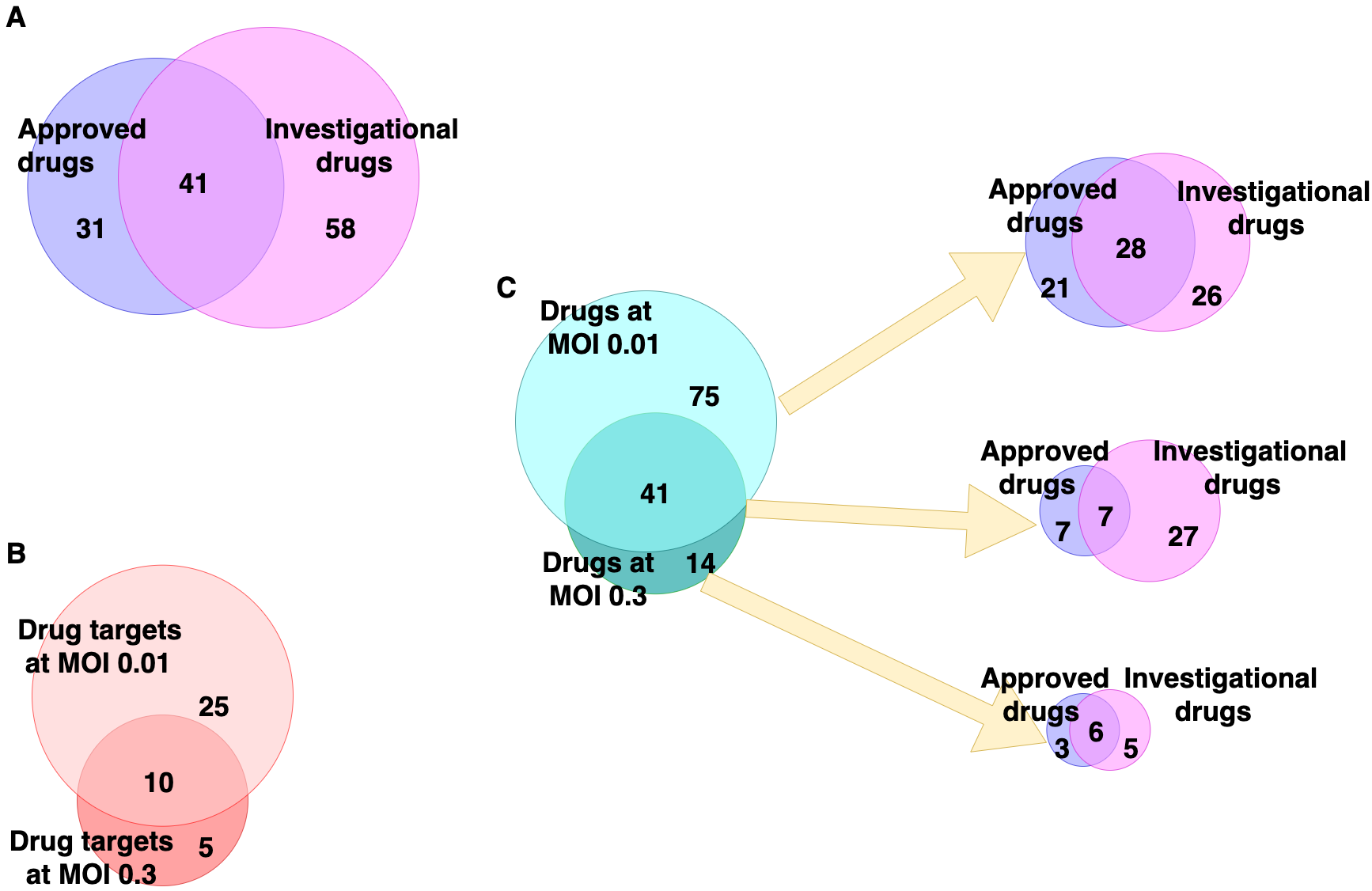}    
    \end{center}
    \caption{\textbf{Drug categories.} \textbf{A}: The overlap between all approved or investigational drugs, \textbf{B}: The overlap between drug-targets in both networks, \textbf{C}: The overlap between approved and investigational drugs, for networks intersection and for set differences }
    \label{figure-categories}
\end{figure}

\begin{figure}[htb]
    \begin{center}
        \includegraphics[width=\textwidth]{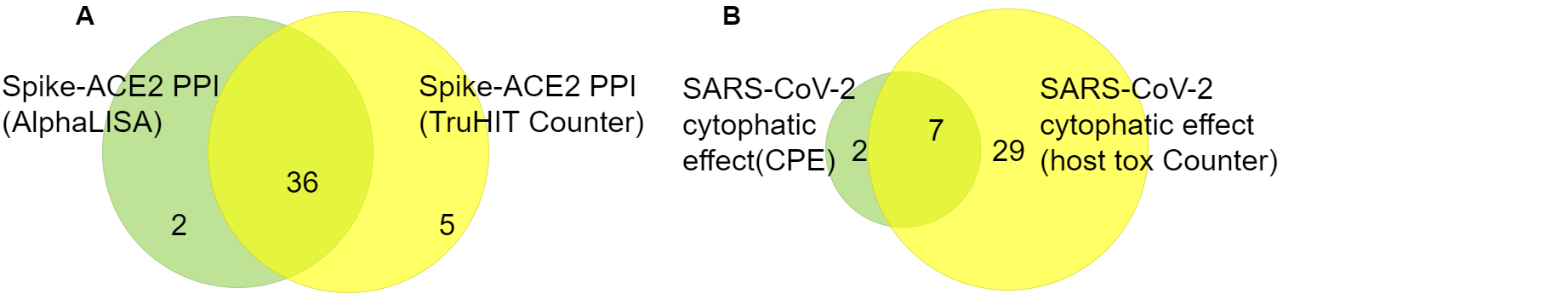}
    \end{center}
    \caption{\textbf{The overlap between drugs active in the assay and in the counter-assay}. \textbf{A}: spike-ACE2 PPI in assay and counter-assay, \textbf{B}: sytophatic effect in assay and counter-assay. \textbf{Color code}: green for the drugs active in the assay, yellow for the drugs active in the counter-assay.}
    \label{figure-overlap}
\end{figure}

\begin{figure}[htb]
    \begin{center}
        \includegraphics[width=0.6\textwidth]{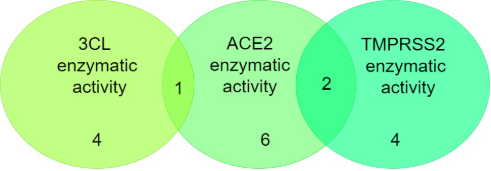}
    \end{center}
    \caption{\textbf{The effect on enzymatic activity}: \textbf{A} - Euler diagram on enzymatic activity, \textbf{B} - 3CLpro enzymatic activity, \textbf{C} - TMPRSS2 enzymatic activity, \textbf{D} - ACE2 enzymatic activity.}
    \label{figure-enzymatic-activity}
\end{figure}

\begin{figure}[htb]
    \begin{center}
        \includegraphics[width=\textwidth]{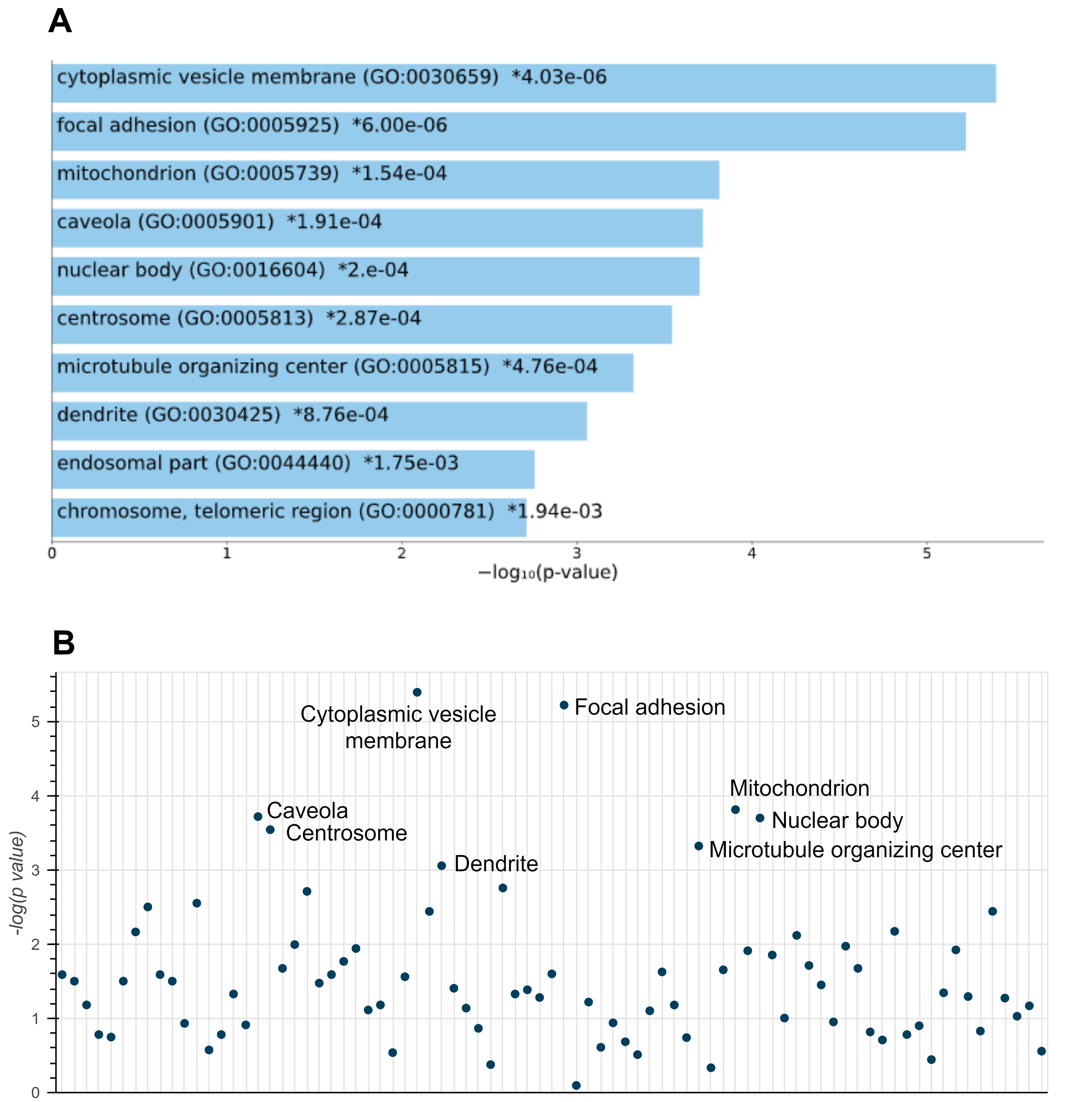}    
    \end{center}
    \caption{\textbf{Localization at the cell level} \cite{clarke2021appyters}, \cite{kuleshov2016enrichr}. \textbf{A}: GO Cellular Component 2018, \textbf{B}: Manhattan plot.}
    \label{figure-localization}
\end{figure}

\begin{figure}[htb]
    \begin{center}
        \includegraphics[height=0.7\textheight]{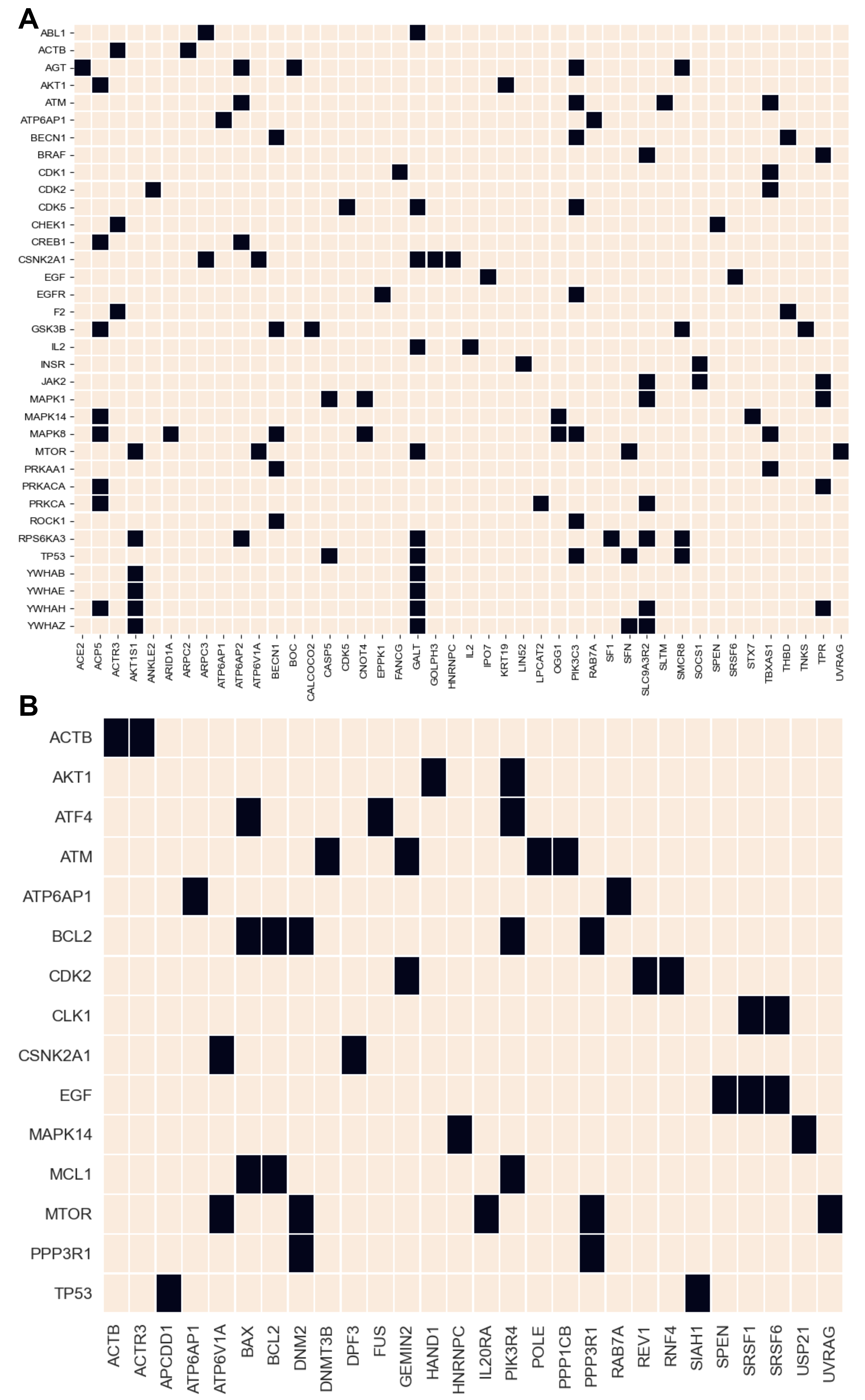}
    \end{center}
    \caption{\textbf{Correspondence between drug-target and SARS-CoV-2 host factors}: \textbf{A} MOI 0.01, \textbf{B} MOI 0.3. On the horizontal axis: the top 200 SARS-CoV-2 host factors (\cite{daniloski2021identification}) included in the interaction network. On the vertical axis: the drug targets included in the interaction network. A black rectangle marks a drug target identified by our analysis to control the corresponding host factor.}
   \label{figure-correspondence-drug-target-essential}
\end{figure}

\end{document}